\documentclass[11pt]{elsarticle}

\usepackage[top=0.1in, bottom=0.9in, left=1in, right=1.2in]{geometry}

\usepackage{graphicx}
\usepackage{subfigure}

\usepackage{amssymb}
\usepackage{amsthm}
\usepackage{amsmath}
\usepackage{amsfonts}

\usepackage[ruled,vlined]{algorithm2e}
\usepackage{mathrsfs}
\usepackage{color}
\usepackage{bm}
\usepackage{arydshln}

\usepackage{hyperref}

\usepackage{fancyhdr}

\usepackage{float}

\date{}
\newcommand{\C} {\mathbb{C}}
\newcommand{\R} {\mathbb{R}}

\newcommand{\InProd}[2] {\left\langle {#1}, {#2} \right\rangle}
\newcommand{\Norm}[2] {\| #1 \|_{#2}}
\newcommand{\Abs}[1] {\left| #1 \right|}

\newcommand{\Define} {:=}

\newcommand{\Projection}[1] {\mathbb{P}{#1}}

\newcommand{\Desired}[1] {\mathbb{D}{(#1)}}

\newcommand{\Threshold}[2] {\mathbb{T}_{#1}{(#2)}} 

\graphicspath{{Figures/}}

\newtheorem{theorem}{Theorem}
\newtheorem{lemma}[theorem]{Lemma}
\newtheorem{proposition}[theorem]{Proposition}
\newtheorem{corollary}[theorem]{Corollary}

\newtheorem{definition}[theorem]{Definition}

\newdefinition{remark}[theorem]{Remark}

\begin{document}

\begin{frontmatter}



\title{Fast thresholding algorithms with feedbacks for sparse signal recovery}


\author[RUC,SFSU]{Shidong Li\fnref{fn1}}
\ead{shidong@sfsu.edu}
\author[RUC]{Yulong Liu\fnref{fn2}}
\ead{yulong3.liu@gmail.com}
\author[RUC]{Tiebin Mi}
\ead{mitiebin@gmail.com}

\fntext[fn1]{S. Li is partially supported by the xxx}
\fntext[fn2]{Y. Liu is a visitor at Renmin University of China.}

\address[RUC]{School of Information, Renmin University of China, Beijing, 100872, China}

\address[SFSU]{Department of Mathematics, San Francisco State University, San Francisco, CA 94132, USA}

\begin{abstract}
We provide another framework of iterative algorithms based on thresholding, feedback and null space tuning for sparse signal recovery arising in sparse representations and compressed sensing. Several thresholding algorithms with various feedbacks are derived, which are seen as exceedingly effective and fast. Convergence results are also provided.  The core algorithm is shown to converge in finite many steps under a (preconditioned) restricted isometry condition. The algorithms are seen as particularly effective for large scale problems.  Numerical studies about the effectiveness and the speed of the algorithms are also presented.
\end{abstract}


\end{frontmatter}



\section{Introduction}
A basic underdetermined linear inverse problem is about the solution to the system of linear equations
\begin{equation}\label{E:Ax_b}
A x = b,
\end{equation}
where $A \in \R^{n \times N}$ ($n \ll N$) and $b \in \R^n$ are known. In the past few years, sparsity constraint has been a popular regularization approach toward the solution of such inverse problems.  The problems of sparse representation and compressed sensing are typical examples.

The goal of sparse representation is to approximate a signal $b$ by a linear combination of the least number of elementary signals drawn from a dictionary $A$. Equivalently, we are to find the sparsest coefficient $x$ such that $Ax=b$. In compressed sensing, signals are assumed to be sparse in some transform domain. The purpose is to recover the coefficient $x$ (and the signal) from a surprisingly small number of linear measurements $Ax=b$. Evidently, the central theme in compressed sensing is also to find sparse solutions to underdetermined linear systems.

With the sparsity constraint, the basic problem here involves finding the sparsest solutions satisfying the linear equations. In other words, one wishes to solve an $\ell_0$-minimization problem
\begin{equation*}
(P_0) \qquad \min_{x \in \R^N} \Norm{x}{0}  \quad \text{s.t.} \quad Ax = b,
\end{equation*}
where $\Norm{\cdot}{0}$ is a quasi-norm standing for the number of the nonzero entries.

$(P_0)$ is clearly combinatorial in nature. It is NP-hard in general \cite{A_Natranjan_SparseApproximateSolutions}.  The renowned advances in this area lie fundamentally in the replacement of $(P_0)$ with a convex relaxation
\begin{equation*}
(P_1) \qquad \min_{x \in \R^N} \Norm{x}{1}  \quad \text{s.t.} \quad Ax = b.
\end{equation*}
See also a series of articles dealing with the equivalence between $(P_0)$ and $(P_1)$, e.g., \cite{A_Candes_CompressiveSampling, A_Candes_RobustUncertaintyPrinciples, A_Donoho_OptimallySparseRepresentation, A_Donoho_StableRecoveryInThePresenceNoise, A_Donoho_UncertaintyPrinciplesIdealAtomicDecomposition, A_Elad_GeneralizedUncertaintyPrinciple, A_Fuchs_SparseRepresentationsArbitraryRedundantBases, A_Gribonval_SparseRepresentationsUnionsBases}. Evidently, $(P_1)$ can be solved accurately by interior-point methods \cite{A_Chen_AtomicDecompositionByBasisPursuit, A_InteriorPointMethodLargeScale_l1_RegularizedLeastSquares} and a number of other different methods.

Among others, ``greedy algorithms'' are another class of popular means of finding sparse solutions.  Two typical representative approaches are Matching Pursuit (MP) and Orthogonal Matching Pursuit (OMP),  e.g., \cite{A_Mallat_MatchingPursuits, A_Needell_UniformUncertaintyPrinciple_RegularizedOrthogonalMatchingPursuit, A_Tropp_GreedIsGood}. In addition, a number of variants of the greedy pursuit algorithms have also been proposed by various authors, e.g., stagewise orthogonal matching pursuit (StOMP) \cite{A_Donoho_StOMP}, compressive sampling matching pursuit (CoSaMP) \cite{A_Needell_CoSaMP} and subspace pursuit (SP) \cite{A_Dai_SubspacePursuit}, etc.

A third class of algorithms for sparse solutions to underdetermined linear inverse problems are iterative thresholding/shrinkage algorithms, which are known for their simplicity and the ease in implementations. Most iterative thresholding/shrinkage algorithms are motivated by minimizing a cost function, which combines a quadratic error term with a sparsity-promoting regularization term, i.e.,
\[
\min_{x \in \R^N} \frac{1}{2} \Norm{Ax - b}{2}^2 + \lambda \Norm{x}{1}.
\]
Various iterative hard/soft thresholding algorithms \cite{A_Dias_NewTwIST, A_Bredies_IHTSparsity, A_Bredies_LinearConvergenceIST, A_Candes_SignalRecoveryFromRandomProjections, A_Daubechies_IterativeThresholdingAlgorithmLinearInverseProblems, A_Fornasier_IterativeThresholdingAlgorithms}, gradient-descent methods \cite{A_Beck_FastIterativeShrinkageThresholdingAlgorithm, A_Figueiredo_GradientProjection, A_Wright_SparseReconstructionSeparableApproximation}, and Bregman iterations \cite{A_Cai_LinearizedBregmanIteration, A_Yin_BregmanIterativeAlgorithmsCS} are representatives. Among this class of works, there is one proposed by Blumensath and Davies \cite{A_Blumensath_IterativeThresholdingSparseApproximations}, in which the $\ell_1$-regularization term is replaced by $\lambda \Norm{x}{0}$. An iterative hard thresholding (IHT) algorithm within the majorization minorization (MM) framework is analyzed \cite{B_Lange_Optimization}. It was also shown that IHT converges to a local minimum of the $\ell_0$-regularized cost function under some conditions.

In \cite{A_Maleki_OptimallyTunedIterativeReconstructionAlgorithms}, Donoho and Maliki combines exact solution to small linear system with
thresholding before and after the solution to derive a more sophisticated scheme, named two-stage thresholding (TST). Very recently, Foucart has proposed a hard thresholding pursuit (HTP) algorithm \cite{A_Foucart_HardThresholdingPursuit}. In essence, HTP can be regarded as a hybrid of IHT and CoSaMP.

In this article, a class of algorithms combining thresholding, feedbacks and null space tuning is proposed to find sparse solutions. The proposed algorithms are brought into a concise framework of {\em null space tuning} (NST). It turns out that the mechanism of NST improves the performance of the algorithms significantly. Several sparsity enhancing  operators are incorporated into the NST to develop various algorithms.  As examples, several specific algorithms are provided which are rather effective in terms of the ``recoverability'' to a larger number of non-zero components/coefficients. These algorithms are shown to be exceedingly fast. Some results about the theoretical performance are also presented. Among this class, two representative algorithms are shown to converge under commonly known conditions.

The organization of this article is as follows. A brief description of the common framework of null space tuning is given in Section~\ref{S:NST}. The core algorithm, null space tuning with hard thresholding and feedback (NST+HT+FB), is introduced in Section~\ref{S:NST_HT_FB}. In Section~\ref{S:OtherChoices}, we present two other algorithms possessing the feedback nature, along with a brief study of the computational issues of the NST based algorithms. Section~\ref{S:Convergence} is dedicated to the theoretical convergence studies of the NST+HT+FB algorithms. We show that the algorithm allows stable recovery of sparse vectors if the measurement matrix satisfies commonly known conditions.  Extensive numerical tests are presented in Section~\ref{S:NumericalExamples} to justify the advantages of the algorithms in practice.

\section{A common framework of the approximation and null space tuning algorithms}\label{S:NST}
Assume that $A$ has full (row) rank and there exists a desired vector $x$ such that $A x = b$. We propose the following iterative framework of the {\em approximation} and \emph{null space tuning} (NST) algorithms
\[
\text{(NST)} \qquad
\left\{
  \begin{array}{ll}
    u^k = \Desired{x^k},  \\
    x^{k+1} = x^k + \Projection{(u^k - x^k)}.
  \end{array}
\right.
\]
Here $\Desired{x^k}$ approximates the desired solution by various principles, and, $\Projection{} \Define I - A^* (A A^*)^{-1} A$ is the orthogonal projection onto $\ker{A}$. The feasibility of $x^0$ is assumed, which guarantees that the sequence $\{ x^k \}$ are all feasible. Obviously, we hope that $u^k \to x$ as $k$ increases.

Due to the feasibility of the sequence $\{ x^k \}$, the NST step can be rewritten as
\begin{equation}\label{E:EquivalentProjection}
\begin{aligned}
x^{k+1}
& = x^k + \Projection{(u^k - x^k)} \\
& = x^k + \left[ I - A^* (A A^*)^{-1} A \right] (u^k - x^k) \\
& = u^k + A^* (A A^*)^{-1} (b - A u^k),
\end{aligned}
\end{equation}
which indicates that $x^{k+1}-u^k$ is perpendicular to the hyperplane $\{ x : A x = b \}$. Therefore, $x^{k+1}$ is the orthogonal projection of $u^k$ onto the feasible set.

In the NST procedure, $\Projection{}$ (or $A^* (A A^*)^{-1}$) can be computed off-line as its appearance does not change during iterations. For very large scale problems, it is very useful to build the two matrices based on the structure of $A$. Oftentimes, the fast Fourier transforms, the wavelet transforms, etc., can be used to facilitate the construction of $\mathbb P$. Using the structure of $A$, the computation may require substantially less memory.

Evidently, the NST framework places much emphasis on the role of the projection. The choice of the approximation operator $\mathbb D$ is definitely a separate topic itself. Different applications call for different operators. As a special case of the NST framework, if $\mathbb D$ is set as a projection onto a convex set (which is not what we suggest here), then the associated NST procedure will clearly be one instance of the POCS (projection onto convex sets) method.

Clearly, with the sparsity constraint, the fundamental of the approximation operator $\mathbb D$ lies in enhancing the sparsity and the feasibility. We would then derive the associated iterative algorithms. In this work, the choices of $\mathbb D$ are not projections onto convex sets. Convergence proofs are also provided, which are anything but trivial since these algorithms are not part of the POCS family.

Specifically, three algorithms will be discussed in this article. These methods have the following appearances ordered in their importance relative to this article. For simplicity, we denote by $T_k$ the index set corresponding to the most $s$ significant entries of $x^k$, by $T_k^c$ the complement set of $T_k$ in $\{1, 2, \cdots, N\}$, and by $A_{T_k}$ the submatrix consisting of columns of $A$ indexed by $T_k$, respectively. We define ${\mathbb T}_s$ the {\em hard thresholding} operator which keeps the largest $s$ entries (in magnitude) and sets all the others to zeros.

\begin{itemize}
\item NST + hard-thresholding + feedback (NST+HT+FB)
\[
\left\{
    \begin{array}{ll}
        u^k_{T_k} = x^k_{T_k} + (A_{T_k}^* A_{T_k})^{-1} A_{T_k}^* A_{T_k^c} x^k_{T_k^c}, \\
        u^k_{T_k^c} = 0, \\
        x^{k+1} = x^k + \Projection{(u^k - x^k)}
    \end{array}
\right.
\]

\item NST + hard-thresholding + suboptimal feedback (NST+HT+subFB)
\[
\left\{
  \begin{array}{ll}
    u^k_{T_k} =  x^k_{T_k} + \lambda^k A_{T_k}^*  A_{T_k^c} x^k_{T_k^c}, \\
    u^k_{T_k^c} = 0, \\
    x^{k+1} = x^k + \Projection{(u^k - x^k)}.
  \end{array}
\right.
\]
Here $\lambda^k \le 1 / \Norm{A_{T_k}^* A_{T_k}}{2}$.

\item NST + stretched hard-thresholding (NST+stretched HT)
\[
\left\{
  \begin{array}{ll}
    u^k = \theta^k \Threshold{s}{x^k}, \\
    x^{k+1} = x^k + \Projection{(u^k - x^k)}.
  \end{array}
\right.
\]
A reasonable choice of $\theta^k$ is $\Norm{b}{1} / \Norm{A_{T_k} x^k_{T_k}}{1}$.

\end{itemize}

\section{Null space tuning with hard thresholding and feedback (NST+HT+FB)}\label{S:NST_HT_FB}
As mentioned earlier, $\mathbb D$ functions fundamentally to approximate the $s$ sparse solution as well as possible. For simplicity, $x^0$ is always set as the least squares solution, i.e.,
$x^0 = A^* (A A^*)^{-1} b$.  Throughout of this article, we shall assume that there exists an $s$ sparse vector $x$ such that $A x = b$.

In the core part of the algorithms, the approximation operator $\mathbb D$ is set as thresholding plus feedback. We begin by commenting that NST+HT+FB is surprisingly efficient and the algorithm converges in finite steps under common assumptions. The convergence is again proven in Section \ref{S:Convergence}.

Since the sequence $\{ x^k \}$ are always feasible in the framework of the NST algorithms, one may split $b$ as
\[
b = A_{T_k} x^k_{T_k} + A_{T_k^c} x^k_{T_k^c}.
\]
In most (if not all) thresholding algorithms, thresholding (hard or soft) is taken by merely keeping the entries of $x^k$ on ${T_k}$, and thereby completely abandons the contribution of $A_{T_k^c} x^k_{T_k^c}$ to the measurement $b$.  Though $x^k_{T^c_k}$ gradually diminishes as $k \to \infty$ (as shown in Section \ref{S:Convergence}), it is not difficult to observe that the contribution of $A_{T_k^c} x^k_{T_k^c}$ to $b$ can be quite significant at initial iterations. Therefore, simple (hard) thresholding alone can be quite infeasible at earlier stages.

We propose an approximation operator $\mathbb D$ that combines the hard thresholding (HT) and a feedback (FB) to enhance the feasibility of $u^k$. This approach is termed the NST+HT+FB algorithm.

The main point is to feed the contribution of $A_{T_k^c} x^k_{T_k^c}$ to $b$ back to $\text{im}(A_{T_k})$, the image of $A_{T_k}$. That is, we require to find a proper $\eta^k$ such that
\[
A_{T_k} \eta^k \approx A_{T_k^c} x^k_{T_k^c}.
\]
A straightforward way is to set
\begin{equation}\label{eqn_feedback}
\eta^k = \arg \min_{\eta} \Norm{A_{T_k} \eta - A_{T_k^c} x^k_{T_k^c}}{2},
\end{equation}
which has the best/least-square solution
\[
\eta^k = (A_{T_k}^* A_{T_k})^{-1} A_{T_k}^* A_{T_k^c} x^k_{T_k^c}.
\]
The NST+HT+FB algorithm is then established as follows
\[
\text{(NST+HT+FB)} \qquad
\left\{
  \begin{array}{ll}
    u^k_{T_k} = x^k_{T_k} + (A_{T_k}^* A_{T_k})^{-1} A_{T_k}^* A_{T_k^c} x^k_{T_k^c}, \\
    u^k_{T_k^c} = 0, \\
    x^{k+1} = x^k + \Projection{(u^k - x^k)} .
  \end{array}
\right.
\]

\begin{figure}[h]
\centering
\includegraphics[width=0.6\columnwidth]{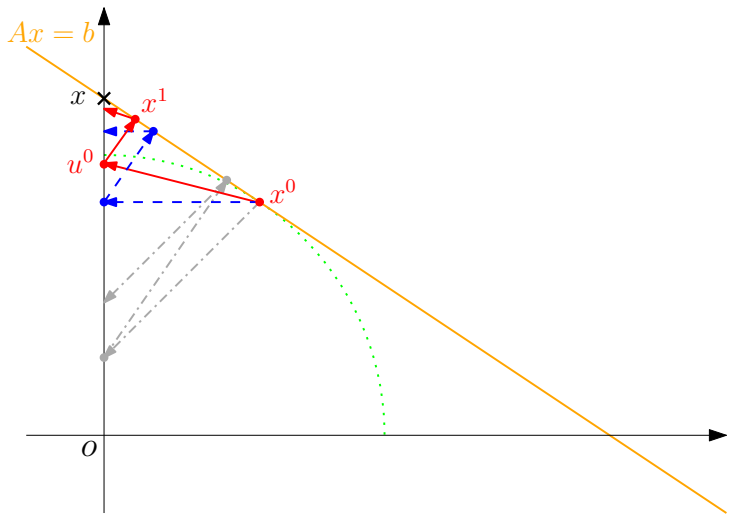}
\caption{Geometric description of the NST based algorithms. The blue (dashed) and the gray (dash dotted) illustrate the trajectories of NST+HT and NST+ST, respectively. The red solid demonstrates the iteration of NST+HT+FB.}
\label{F:NST}
\end{figure}

Figure~\ref{F:NST} depicts the geometric interpretation of the NST based algorithms. Here $n=1$, $N=2$ and $s=1$. The red trajectory is a geometric interpretation of the NST+HT+FB algorithm. The algorithm is deemed converged when $\Norm{A u^k - b}{2}$ is less than some predetermined value, or $\Norm{u^k - u^{k-1}}{2}$ is sufficiently small. The convergence proof is placed in Section \ref{S:Convergence}.  Algorithm~\ref{A:NST_HT_FB} is the pseudo-code of the NST+HT+FB procedure.

\begin{algorithm}\label{A:NST_HT_FB}
\caption{NST+HT+FB}
\KwIn{$A$, $b$, $s$, $x_{o}$, $\epsilon_1$, $\epsilon_2$;}
\KwOut{$u$, $x$;}
$u^{-1} = 0$; $k = 0$; $x^{0} = x_{o}$;

$u^{0} = \Threshold{s}{x^{0}}$;

\While{$\Norm{A u_{k} - b}{2} / \Norm{b}{2} \ge \epsilon_1$ and $\Norm{u^{k}-u^{k-1}}{2} / \Norm{u^{k-1}}{2} \ge \epsilon_2$}
{
$k = k+1$;

$u^{k} = \Threshold{s}{x^k} + (A_{T_k}^* A_{T_k})^{-1} A_{T_k}^* A_{T_k^c} x^k_{T_k^c}$;

$x^{k+1} = x^{k}+\mathbb{P} (u^{k} - x^{k})$;
}
\Return{$u^{k}, x^{k}$};
\end{algorithm}
\medskip

We may also compare the NST+HT+FB with a simpler thresholding-only scheme within the NST framework which we discuss immediately later.  It is this very feedback adjustment that greatly improves the rate of convergence.

In fact, for cases in $\R^2$, NST+HT+FB stops in just 1 iteration. Let us examine the example illustrated in Figure~\ref{F:NST}, where the underdetermined system is $a_1 x_1 + a_2 x_2 = b$ and $a_1 > a_2 > 0$. Clearly, the exact solution is $x = [0 \ \frac{b}{a_2}]^T$ and $x^0 = [\frac{a_1 b}{a_1^2+a_2^2} \ \frac{a_2 b}{a_1^2+a_2^2}]^T$. The equation $a_2 \eta^{0} = a_1 x^{0}_1$ has definitely an exact solution $\eta^0 = a_2^{-1} a_1 x^{0}_1$. This feedback in turn gives rise to $u^0 = [0 \ \frac{b}{a_2}]^T$, which is the exact solution. By the same principle, the readers shall see in Section~\ref{S:Convergence} that the NST+HT+FB algorithm generally converges in finite steps.



As a comparison, let us also understand a bit more about a thresholding-only scheme under the NST framework. One natural choice of the approximation operator $\mathbb D$ is to keep only the largest $s$ entries, which gives rise to the simpler NST+HT algorithm
\[
\text{(NST+HT)} \qquad
\left\{
  \begin{array}{ll}
    u^k = \Threshold{s}{x^k}, \\
    x^{k+1} = x^k + \Projection{(u^k - x^k)}.
  \end{array}
\right.
\]
Clearly,  $u^k$ can be regarded as a projection onto the highly non-convex ``$\ell_0$-ball'' ($\{ x : \Norm{x}{0} \le s \}$) in some sense.

We comment that $\Projection{(u^k - x^k)}$ is the nearest element in $\ker{A}$ to $u^k - x^k$ in the $\ell_2$-norm sense, i.e.,
\[
\Projection{(u^k - x^k)} = \arg \min_{z \in \ker{A}} \Norm{(u^k - x^k) - z}{2}.
\]
Though $u^k - x^k$ cuts out the $(N-s)$ smaller entries of $x^k$ completely,  $\Projection{(u^k - x^k)}$ provides necessary corrections in the least squares sense. This can explain partially why NST+HT enhances the sparsity gradually. The convergence of NST+HT will be discussed in Section~\ref{S:Convergence}, where we show that $\lim_{k \to \infty} u^k = x$ under commonly known conditions.

\begin{remark} \ As an example of the NST framework, Cand\`{e}s and Romberg's work \cite{A_Candes_SignalRecoveryFromRandomProjections} is in fact a null space tuning with {\em soft thresholding} (NST+ST) in an attempt to provide a solution to $(P_1)$.  Soft thresholding approximates a projection onto the convex $\ell_1$-ball.  Therefore, the convergence of their algorithm could fall within the framework POCS.  The NST+HT method, however, is not a POCS algorithm. As we shall show in Section \ref{S:Convergence}, the convergence of NST+HT can not generally follow the threads of POCS convergence arguments. A preconditioned restricted isometry condition is introduced instead.
\end{remark}

The blue (dashed) and the gray (dash dotted) in Figure~\ref{F:NST} illustrate the trajectories of NST+HT and NST+ST, respectively. Clearly, the hard thresholding operator finds $u^k$ by keeping the most significant entry of $x^k$ and the soft thresholding operator finds $u^k$ by shrinking all entries of $x^k$ toward zero. Figure~\ref{F:NST} explains intuitively why NST+HT possesses greater rate of convergence than that of NST+ST in general.  Figure~\ref{F:NST} also provides an intuitive description (in red solid) why the NST+HT+FB tends to converge faster.

The NST+HT algorithm has seemingly some resemblance to the method of iterative hard thresholding (IHT). Immediate discussions about the differences and the connections between NST+HT and IHT are the very next concern.

\subsection*{Differences and connections between NST+HT and IHT}
The iteration of IHT proposed in, e.g., \cite{A_Blumensath_IterativeThresholdingSparseApproximations, A_Blumensath_IHTCompressedSensing} is
\[
\text{(IHT)} \qquad
\left\{
  \begin{array}{ll}
    u^k = \Threshold{s}{x^k}, \\
    x^{k+1} = u^k + A^* (b - A u^k).
  \end{array}
\right.
\]
NST+HT, on the other hand, has an equivalent form
\[
\left\{
  \begin{array}{ll}
    u^k = \Threshold{s}{x^k}, \\
    x^{k+1} = u^k + A^* (A A^*)^{-1} (b - A u^k).
  \end{array}
\right.
\]
Evidently, in a special case where $A$ is a Parseval frame, i.e., $A A^* = I$, NST+HT reduces to IHT.

However, for general measurement matrices, considerable amount of numerical studies demonstrate that NST+HT is far more effective than IHT. More precisely, NST+HT has a much greater ability to recover signals consisting of a much larger number of nonzero components than that of IHT.  These will be illustrated in Section~\ref{S:NumericalExamples}. The key rational of such performance differentiations lies in the null space tuning step.

\subsection*{A point of view from error correction (denosing)}
We may also peek into the contaminated signal situation, and discover that the role of  $(A A^*)^{-1}$ is vital at reducing the noise. That explains why NST+HT outperforms IHT in realistic situations where signals are contaminated by noises, or when signals are approximately sparse. In the following, we assume $x = x^\sharp + v$, where $x^\sharp$ is the ideal $s$ sparse signal and $v$ is the Gaussian white noise.

As analyzed in \cite{A_Blumensath_IterativeThresholdingSparseApproximations}, IHT can be regarded as a majorization minorization (MM) algorithm to the problem
\[
(P_{0,s}) \qquad \min_{x \in \R^N} \Norm{Ax - b}{2} \quad \text{s.t.} \quad \Norm{x}{0} \le s.
\]
Evidently, solving $(P_{0,s})$ is to find the best $s$-term approximation to the original signal from the given measurements.

Let us observe, however, that the cost function in $(P_{0,s})$ is not quite ideal for finding good approximations to contaminated (not exactly sparse) signals. The following is a brief discussion.

Suppose $u$ is the solution to $(P_{0,s})$. The original signal can be written as $x = u + e$, where $e$ is the approximation error (nonzero by the fact that $x$ is not precisely sparse), and the measurement can be written as $A (u + e) = b$. For simplicity, we assume there is no measurement error. By the singular value decomposition, $A = U \Sigma V^*$, where $U$ and $V$ are unitary and $\Sigma$ is the $n\times N$ singular matrix having in the main diagonal singular values $\{\sigma_i\}_{i=1}^n$. It then follows that $\Norm{A e}{2} = \Norm{U \Sigma V^* e}{2} = \Norm{\Sigma V^* e}{2} = \Norm{\Sigma \tilde{e}}{2}$, where $\tilde{e} \Define V^* e$. Let us note that the $i^{th}$ component of $\tilde{e}$ is therefore the approximation error in the direction identified by the $i^{th}$ column of $V$.

Recall that $u$ is the solution to  $(P_{0,s})$. Hence, $\Norm{A u - b}{2} = \Norm{Ae}{2} = \Norm{\Sigma \tilde{e}}{2}$ possesses the least $\ell_2$-norm. That is to say $(P_{0,s})$ is to minimize $\Norm{\Sigma \tilde{e}}{2}$. However, minimizing $\Norm{\Sigma \tilde{e}}{2}$ would penalize more heavily the error components corresponding to the larger singular values.  Hence, the solution to $(P_{0,s})$ would depend on the singular value structure of the measurement matrix $A$, which can be problematic.  Because the original signal is completely independent of the measurement system $A$, the approximation error corresponding to the appropriate solution should be penalized fairly in all (singular vector) directions.

We can now explain why NST+HT tends to perform better by the same MM principle. The null space tuning step can be written further as
\[
\begin{aligned}
x^{k+1}
& = u^k + A^* (A A^*)^{-1} (b - A u^k) \\
& = u^k + A^* (A A^*)^{-\frac{1}{2}} [(A A^*)^{-\frac{1}{2}}b - (A A^*)^{-\frac{1}{2}} A u^k ]
\end{aligned}
\]
Consequently, with an equivalent measurement equation $(A A^*)^{-\frac{1}{2}} A x =(A A^*)^{-\frac{1}{2}}b$ and the measurement matrix
$(A A^*)^{-\frac{1}{2}} A$ , NST+HT may also be regarded as a majorization minorization (MM) algorithm for the solution of the following problem
\[
(P'_{0,s}) \qquad \min_{x \in \R^N} \Norm{(A A^*)^{-\frac{1}{2}} (Ax - b)}{2} \quad \text{s.t.} \quad \Norm{x}{0} \le s.
\]
Similar to the previous analysis, $(P'_{0,s})$ is to minimize $\Norm{(A A^*)^{-\frac{1}{2}} A e}{2}$, where $e = x - u$ is the approximation error, $x$ is the contaminated signal and $u$ is the solution to $(P'_{0,s})$ as well. With $A = U \Sigma V^*$, $(A A^*)^{-\frac{1}{2}} A = U [I \ 0] V^*$, and $\Norm{(A A^*)^{-\frac{1}{2}} A e}{2} = \Norm{U [I \ 0] V^* e}{2} = \Norm{[I \ 0] \tilde{e}}{2}$. Therefore, the approximation error is penalized fairly in all directions identified by the associated singular vectors of the measurement system.

\begin{figure}
\centering
\includegraphics[width=0.5\columnwidth]{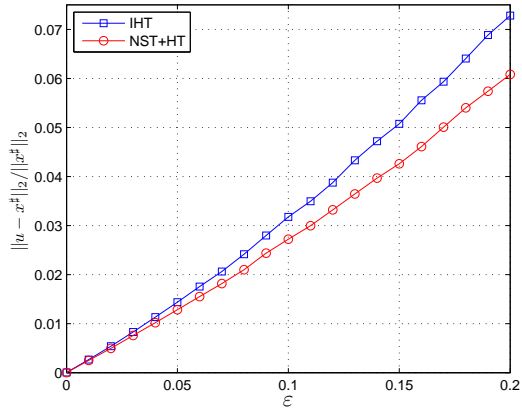}
\caption{Plots of $\Norm{u - x^\sharp}{2}/\Norm{x^\sharp}{2}$ as a function of the noise level $\varepsilon$ for NST+HT and IHT. Here $s=5$, $\Norm{x^\sharp}{2} = 1$ and $\Norm{v}{2} = \epsilon$. NST+HT does have the advantage at recovering contaminated Gaussian sparse vectors over that of IHT.}
\label{F:RelativeErrorIHTNoisySignal}
\end{figure}

This explains why NST+HT functions better in denoising than IHT for general measurement matrices from the viewpoint of MM procedures.  For a quick demonstration, Figure~\ref{F:RelativeErrorIHTNoisySignal} is the numerical example showing this very fact about the error correction differentials. In this experiment, we just set $x = x^{\sharp}+v$. Here $x^{\sharp}$ is the original sparse signal with the unit $\ell_2$-norm and $v$ is the Gaussian white noise with the energy $\epsilon$.

We leave the convergence study to Section \ref{S:Convergence}, and move on to the rest algorithms of the NST family.

\section{Other choices of $\mathbb D$ and computational issues}\label{S:OtherChoices}
In the NST+HT+FB scheme, a matrix inversion $(A_{T_k}^* A_{T_k})^{-1}$ is required, and $T_k$ changes during iterations. The inversion may then need to be computed in every step of the iterations. For large scale problems, such matrix inversion in every step (as in OMP, CoSaMP, SP and HTP) can still be computationally intensive.

We propose here some other choices of the approximation operator $\mathbb D$ having the feedback functionality, but with substantially less computational complexities.

\subsection{NST+HT+suboptimal feedbacks}
Note that the role of $(A_{T_k}^* A_{T_k})^{-1} A_{T_k}^* A_{T_k^c} x^k_{T_k^c}$ is to calculate the feedback of the tail contribution to $b$. As long as $A_{T_k}^* A_{T_k}$ is well conditioned, which is a typical requirement by the  well-known restricted isometry property (RIP) for convergence proofs (Section~\ref{S:Convergence}), $(A_{T_k}^* A_{T_k})^{-1}$ can be approximated by $\lambda^k I$ with $\lambda^k$ being on the order of the spectrum of $(A_{T_k}^* A_{T_k})^{-1}$. Evidently, a natural approximation of the feedback can be simplified to $\lambda^k A_{T_k}^*  A_{T_k^c} x^k_{T_k^c}$. We then reach the first suboptimal scheme
\[
\text{(NST+HT+subFB)} \qquad
\left\{
  \begin{array}{ll}
    u^k_{T_k} = x^k_{T_k} + \lambda^k A_{T_k}^*  A_{T_k^c} x^k_{T_k^c}, \\
    u^k_{T_k^c} = 0, \\
    x^{k+1} = x^k + \Projection{(u^k - x^k)}.
  \end{array}
\right.
\]
Here we choose $\lambda^k \le 1 / \Norm{A_{T_k}^* A_{T_k}}{2}$ for the evident stability consideration of the algorithm.

\subsection{NST+HT with stretching}
Let us recall that the purpose of the feedback is to enhance the feasibility of $u^{k}$. We also observe that, to the most $s$ significant components of $x^k$, the feedback action results in oftentimes the magnification of the components. With this observation, we also propose a stretched/scaled version of NST+HT
\[
\text{(NST+stretchedHT)} \qquad
\left\{
  \begin{array}{ll}
    u^k = \theta^k \Threshold{s}{x^k}, \\
    x^{k+1} = x^k + \Projection{(u^k - x^k)}.
  \end{array}
\right.
\]
Evidently, we hope $A_{T_k} \theta^k x^k_{T_k}$ approximates $b$ better than $A_{T_k} x^k_{T_k}$. A reasonable choice of $\theta^k$ is $\Norm{b}{1} / \Norm{A_{T_k} x^k_{T_k}}{1}$.

\subsection{Sparsity adaptation}
We note that all previous algorithms require the knowledge of the sparsity of the desired solution. This seems wishful in most applications. To make the NST based algorithms more applicable, we may begin with a conservative estimation $s_0$ of the sparsity of the real solution, and increase the sparsity by $s'$ gradually. Taking NST+HT as an example, Algorithm~\ref{A:AdaptiveNST} describes the pseudo-code of the adaptive NST+HT algorithm.

\begin{algorithm}\label{A:AdaptiveNST}
\caption{Adaptive NST+HT algorithm}
\KwIn{$A$, $b$, $s_0$, $s'$, $s_1$, $\epsilon_1$, $\epsilon_2$;}
\KwOut{$u$, $x$;}

$u_{o} = 0$; $s = s_0$;

$x_{o} = A^* (A A^*)^{-1} b$;

$(u_{n},x) = \text{(NST+HT)}(A,b,s,x_{o},\epsilon_1,\epsilon_2)$;

\While{$\Norm{A u_{n} - b}{2} / \Norm{b}{2} \ge \epsilon_1$ and $\Norm{u_{n}-u_{o}}{2} / \Norm{u_{o}}{2} \ge \epsilon_2$ and $s <= s_1$}
{
$u_{o} = u_{n}$;

$s = s+s'$; $x_{o} = x$;

$(u_{n},x) = \text{(NST+HT)}(A,b,s,x_{o},\epsilon_1,\epsilon_2)$;
}

\Return{$u_{n}$, $x$};

\end{algorithm}

\subsection{Computational complexity per iteration}
We now study the computational issues about the NST based algorithms briefly.

For the null space tuning step, $A^* (A A^*)^{-1}$ does not change the appearance during iterations. Consequently, if the inversion $(A A^*)^{-1}$ is calculated off-line and stored in the memory, the NST step requires $n(N+n+s)$ multiplications. As mentioned earlier, for very large scale problems, it is necessary to exploit the structure of the matrices. The computational complexity of the null space tuning step may be reduced substantially to $O(N \log n)$.

The computational complexity of the approximation operator varies greatly for different algorithms. For NST+HT+FB, solving the least squares problem \eqref{eqn_feedback} by Cholesky factorization requires $\frac{1}{2} n s^2 +\frac{1}{6}s^3 + O(n s)$ multiplications \cite{B_NumericalMethodLeastSquaresProblems}. For NST+HT+subFB, the typical number of multiplicative operations of calculating $A_{T_k}^*  A_{T_k^c} x^k_{T_k^c}$ is $n N$. Furthermore, if the Gramian matrix $A^* A$ is stored in the memory, the number is reduced to $s (N-s)$. For NST+stretchedHT, to calculate the scalar $\theta^k$ requires $s n$ multiplications and a division.

To sum up, when $(A A^*)^{-1}$ is stored in the memory and $n$ is on the same order of $N$, NST+HT+FB has a computational complexity at each iteration $\frac{1}{2} n s^2 + \frac{1}{6} s^3 + O(n N)$ for general matrices. For the other three NST based algorithms, the computational complexity is $O(n N)$.

\section{Convergence}\label{S:Convergence}
In this section, theoretical performances of the NST+HT and the NST+HT+FB algorithms are presented.  Convergence results are first established when the sequence $\{ T_k \}$ are the same.  This is followed by clarifications of the connection between NST+HT and NST+HT+FB. Before continuing, some preparations are in the order.

\begin{proposition}[Weyl's inequality \cite{B_MatrixAnalysis}]
Let $X, Y \in \C^{n \times n}$ be Hermitian matrices and let the eigenvalues $\lambda_{i}(X)$, $\lambda_{i}(Y)$ and $\lambda_{i}(X+Y)$ be arranged in descending orders. Then for each $k = 1,2,\ldots,n$,
\[
\lambda_{k}(X) + \lambda_{n}(Y) \le \lambda_{k}(X+Y) \le \lambda_{k}(X) + \lambda_{1}(Y).
\]
\end{proposition}

\begin{lemma}\label{T:NormSubMatrices}
Let $A \in \R^{n \times N}$ be a Parseval frame and $T \subset \{ 1, \ldots, N \}$ with $\Abs{T} < n$. If $A_{T^c} A_{T^c}^*$ is invertible, then $\Norm{A_{T} A_{T}^*}{2} < 1$ and $\Norm{A_{T^c} A_{T^c}^*}{2} = 1$.
\end{lemma}

\begin{proof}
Because $A_{T} A_{T}^*$ and $A_{T^c} A_{T^c}^*$ are all non-negative definite, $\Norm{A_{T} A_{T}^*}{2} = \lambda_{1}(A_{T} A_{T}^*)$ and $\Norm{A_{T^c} A_{T^c}^*}{2} = \lambda_{1} (A_{T^c} A_{T^c}^*)$, respectively. We then require to prove $\lambda_{1}(A_{T} A_{T}^*) < 1$ and $ \lambda_{1} (A_{T^c} A_{T^c}^*) = 1$.

Since $A$ is a Parseval frame, $A A^* = A_{T} A_{T}^* + A_{T^c} A_{T^c}^* = I$, which means $\lambda_i(A_{T} A_{T}^* + A_{T^c} A_{T^c}^*) = 1$, for $i = 1, \ldots, n$. According to Weyl's inequality,
\[
\lambda_{1} (A_{T} A_{T}^*) + \lambda_{n} (A_{T^c} A_{T^c}^*) \le \lambda_{1}(A_{T} A_{T}^* + A_{T^c} A_{T^c}^*) = 1.
\]
Furthermore, $A_{T^c} A_{T^c}^*$ is invertible, implying $\lambda_n (A_{T^c} A_{T^c}^*) > 0$. Then $\lambda_{1}(A_{T} A_{T}^*) < 1$ is followed.

For the second claim, we derive
\[
\lambda_{n} (A_{T} A_{T}^*) + \lambda_{1} (A_{T^c} A_{T^c}^*) \ge \lambda_{n} (A_{T} A_{T}^* + A_{T^c} A_{T^c}^*) = 1
\]
and
\[
\lambda_{1} (A_{T^c} A_{T^c}^*) + \lambda_{n} (A_{T} A_{T}^*) \le \lambda_{1} (A_{T^c} A_{T^c}^* + A_{T} A_{T}^*) = 1.
\]
These two inequalities lead to $\lambda_{1} (A_{T^c} A_{T^c}^*) + \lambda_{n} (A_{T} A_{T}^*) = 1$. Since $\Abs{T} < n$, $\lambda_{n} (A_{T} A_{T}^*) = 0$. Hence, $\lambda_{1} (A_{T^c} A_{T^c}^*) = 1$ as asserted.
\end{proof}

\begin{theorem}\label{T:NST_HT_FixSupport}
Let $A$ be a Parseval frame. Suppose $A_{T}^* A_{T}$ and $A_{T^c} A_{T^c}^*$ are invertible with $T \subset \{ 1, \ldots, N \}$ and $\Abs{T} = s$. For NST+HT, if $T_{j} = T_{j+1} = \ldots = T$ for some integer $j$, then $\lim_{k \to \infty} x^k = x^\natural$. Here $x^\natural$ has the expression
\[
x^\natural_T = x^{j}_T + (A_T^* A_T)^{-1} A_T^* A_{T^c} x_{T^c}^{j}
\]
and
\[
x^\natural_{T^c} = A_{T^c}^* \left( I - A_T (A_T^* A_T)^{-1} A_T^* \right) A_{T^c} x_{T^c}^{j}.
\]
\end{theorem}

\begin{proof}
Without loss of generality, we assume $T = \{ 1, \ldots, s \}$ and $j = 0$. We then require to prove
\[
x^\natural_T = x^{0}_T + (A_T^* A_T)^{-1} A_T^* A_{T^c} x_{T^c}^{0}
\]
and
\[
x^\natural_{T^c} = A_{T^c}^* \left( I - A_T (A_T^* A_T)^{-1} A_T^* \right) A_{T^c} x_{T^c}^{0}.
\]

When $A$ is a Parseval frame and $T_0 = T_1 = \ldots = T$, the second step of NST+HT gives rise to
\[
\begin{aligned}
\begin{bmatrix}
x_{T}^{k+1} \\
x_{T^c}^{k+1} \\
\end{bmatrix}
& =
\begin{bmatrix}
x_{T}^{k} \\
x_{T^c}^{k} \\
\end{bmatrix}
-
\Projection{\left(
\begin{bmatrix}
0 \\
x_{T^{c}}^{k} \\
\end{bmatrix} \right) } \\
& =
\begin{bmatrix}
x_T^{k} \\
x_{T^c}^{k} \\
\end{bmatrix}
- \left( I -
\begin{bmatrix}
A_T^* \\
A_{T^c}^* \\
\end{bmatrix}
\begin{bmatrix}
A_{T} & A_{T^c}
\end{bmatrix}
\right)
\begin{bmatrix}
0 \\
x_{T^c}^{k} \\
\end{bmatrix} \\
& =
\begin{bmatrix}
x_{T}^{k} + A_T^* A_{T^c} x_{T^c}^{k} \\
A_{T^c}^* A_{T^c} x_{T^c}^{k}
\end{bmatrix} .
\end{aligned}
\]
Therefore,
\[
\begin{bmatrix}
x_{T}^{k+1} \\
x_{T^c}^{k+1} \\
\end{bmatrix}
=
\begin{bmatrix}
x_T^{0} \\
0 \\
\end{bmatrix}
+
\begin{bmatrix}
A_T^* + A_T^* (A_{T^c} A_{T^c}^*) + \cdots + A_T^* (A_{T^c} A_{T^c}^*)^{k} \\
A_{T^c}^* (A_{T^c} A_{T^c}^*)^{k} \\
\end{bmatrix} A_{T^c} x_{T^c}^{0} .
\]
We now establish the required statements
\begin{equation}\label{E:NeumannSeries}
\lim_{k \to \infty} \left( A_T^* + A_T^* (A_{T^c} A_{T^c}^*) + \cdots + A_T^* (A_{T^c} A_{T^c}^*)^{k} \right) = (A_T^* A_T)^{-1} A_T^*
\end{equation}
and
\[
\lim_{k \to \infty} (A_{T^c} A_{T^c}^*)^{k} = I - A_T (A_T^* A_T)^{-1} A_T^*.
\]

Since $A_{T} A_{T}^*$ and $A_{T}^* A_{T}$ are all non-negative definite, $\Norm{A_{T} A_{T}^*}{2} = \lambda_{1} (A_{T} A_{T}^*) = \lambda_{1} (A_{T}^* A_{T}) = \Norm{A_{T}^* A_{T}}{2}$. Lemma~\ref{T:NormSubMatrices} implies, when $A_{T^c} A_{T^c}^*$ is invertible, $\Norm{A_{T} A_{T}^*}{2} < 1$. Together with the invertibility of $A_{T}^* A_{T}$, we then derive $0 < I - A_{T}^* A_{T} < I$.

Furthermore, since $A_{T} A_{T}^* + A_{T^c} A_{T^c}^* = A A^* = I$, for each integer $k > 0$,
\[
\begin{aligned}
A_T^* (A_{T^c} A_{T^c}^*)^{k}
= & A_T^* (I - A_{T} A_{T}^*)^{k} \\
= & A_T^* \sum_{l=0}^{k} \binom{k}{l} (- A_{T} A_{T}^*)^{l} \\
= & \sum_{l=0}^{k} \binom{k}{l} (- A_{T}^* A_{T})^{l} A_T^* \\
= & (I - A_{T}^* A_{T})^{k} A_T^* .
\end{aligned}
\]
Consequently, utilizing the Neumann series,
\[
\begin{aligned}
& \lim_{k \to \infty} \left( A_T^* + A_T^* (A_{T^c} A_{T^c}^*) + \cdots + A_T^* (A_{T^c} A_{T^c}^*)^{k} \right) \\
= & \lim_{k \to \infty} \left( A_T^* + (I - A_{T}^* A_{T}) A_T^* + \cdots + (I - A_{T}^* A_{T})^{k} A_T^* \right) \\
= & \left( \lim_{k \to \infty} \sum_{l=0}^{k} (I - A_{T}^* A_{T})^{l} \right) A_{T}^* \\
= & (A_{T}^* A_{T})^{-1} A_{T}^* .
\end{aligned}
\]
The first claim is then followed.

Similarly, for integer $k > 0$,
\[
\begin{aligned}
(A_{T^c} A_{T^c}^*)^{k}
= & I - (I - A_{T^c} A_{T^c}^*) \sum_{l=0}^{k-1} (A_{T^c} A_{T^c}^*)^{l} \\
= & I - A_{T} A_{T}^* \sum_{l=0}^{k-1} (I - A_{T} A_{T}^*)^{l} \\
= & I - A_{T} A_{T}^* \sum_{l=0}^{k-1} \sum_{m=0}^{l} \binom{l}{m} (- A_{T} A_{T}^*)^{m} \\
= & I - A_{T} \left( \sum_{l=0}^{k-1} \sum_{m=0}^{l} \binom{l}{m} (- A_{T}^* A_{T})^{m} \right) A_{T}^* \\
= & I - A_{T} \left( \sum_{l=0}^{k-1} (I - A_{T}^* A_{T})^{l} \right) A_{T}^* .
\end{aligned}
\]
Exploiting the Neumann series again leads to
\[
\begin{aligned}
\lim_{k \to \infty} (A_{T^c} A_{T^c}^*)^{k}
= & I - A_{T} \left( \lim_{k \to \infty} \sum_{l=0}^{k-1} (I - A_{T}^* A_{T})^{l} \right) A_{T}^* \\
= & I - A_{T} (A_{T}^* A_{T})^{-1} A_{T}^* .
\end{aligned}
\]
The second claim is then followed.
\end{proof}

\begin{remark}
In real applications, the invertibility of $A_{T}^* A_{T}$ and $A_{T^c} A_{T^c}^*$ is a weak/minor assumption. It's noteworthy to point out that, to prove \eqref{E:NeumannSeries}, the Neumann series can not be applied directly to
\[
\lim_{k \to \infty} A_T^* \left( I + A_{T^c} A_{T^c}^* + \cdots + (A_{T^c} A_{T^c}^*)^{k} \right).
\]
This is because, when $\Abs{T} < n$ and $A_{T^c} A_{T^c}^*$ is invertible, $\Norm{A_{T^c} A_{T^c}^*}{2} = 1$, as demonstrated in Lemma~\ref{T:NormSubMatrices}.
\end{remark}

\begin{theorem}\label{T:NST_HT_FB_FixSupport}
Suppose $A_{T}^* A_{T}$ is invertible with $T \subset \{ 1, \ldots, N \}$ and $\Abs{T} = s$. For NST+HT+FB, if $T_{j} = T_{j+1} = \ldots = T$ for some integer $j$, then $x^{j+1} = x^{j+2}= \ldots = x^\ddag$. Here $x^\ddag$ has the expression
\[
x^\ddag_{T} = x_{T}^{j} + \left[ (A_{T}^* A_{T})^{-1} A_{T}^* + A_{T}^* ( A A^* )^{-1} \left( I - A_T (A_{T}^* A_{T})^{-1} A_{T}^* \right) \right] A_{T^c} x_{T^c}^{j}
\]
and
\[
x^\ddag_{T^c} = A_{T^c}^* \left( A A^* \right)^{-1} \left( I - A_T (A_{T}^* A_{T})^{-1} A_{T}^* \right) A_{T^c} x_{T^c}^{j} .
\]
\end{theorem}

\begin{proof}
Without loss of generality, we assume $T = \{ 1, \ldots, s \}$. Then the second step of NST+HT+FB gives rise to
\[
\begin{aligned}
\begin{bmatrix}
x_{T}^{j+1} \\
x_{T^c}^{j+1} \\
\end{bmatrix}
= &
\begin{bmatrix}
x_T^{j} \\
x_{T^c}^{j} \\
\end{bmatrix}
+
\Projection{ \left(
\begin{bmatrix}
x^j_T + (A_T^* A_T)^{-1} A_T^* A_{T^c} x^j_{T^c} \\
0 \\
\end{bmatrix}
-
\begin{bmatrix}
x_T^{j} \\
x_{T^c}^{j} \\
\end{bmatrix}
\right)} \\
= &
\begin{bmatrix}
x_T^{j} \\
x_{T^c}^{j} \\
\end{bmatrix}
+ \left( I - A^*
\left(
A A^* \right)^{-1}
\begin{bmatrix}
A_{T} & A_{T^c}
\end{bmatrix}
\right)
\begin{bmatrix}
(A_{T}^* A_{T})^{-1} A_{T}^* A_{T^c}  \\
-I \\
\end{bmatrix}
x_{T^c}^{j} \\
= &
\begin{bmatrix}
x_{T}^{j} + (A_{T}^* A_{T})^{-1} A_{T}^* A_{T^c} x_{T^c}^{j} \\
0
\end{bmatrix} + \\
&
\begin{bmatrix}
A_{T}^* \\
A_{T^c}^* \\
\end{bmatrix}
\left( A A^* \right)^{-1}
\left( I - A_T (A_{T}^* A_{T})^{-1} A_{T}^* \right) A_{T^c} x_{T^c}^{j} .
\end{aligned}
\]
That is $x^{j+1} = x^\ddag$.

We next prove $\Projection{\left( u^{j+1} - x^{j+1} \right)} = 0$, which implies $x^{j+1} = x^{j+2} = \ldots$. For simplicity, define $\mathbb{Q} \Define I - A_T (A_{T}^* A_{T})^{-1} A_{T}^*$. Evidently, $A_T^* \mathbb{Q} = 0$ and $\mathbb{Q} A_T = 0$.

Since $x^{j+1}_{T^c} = A_{T^c}^* ( A A^* )^{-1} \mathbb{Q} A_{T^c} x_{T^c}^{j}$,
\[
\begin{aligned}
u^{j+1} - x^{j+1}
= &
\begin{bmatrix}
(A_{T}^* A_{T})^{-1} A_{T}^* A_{T^c} \\
-I \\
\end{bmatrix}
x_{T^{c}}^{j+1} \\
= &
\begin{bmatrix}
(A_{T}^* A_{T})^{-1} A_{T}^* A_{T^c} \\
- I \\
\end{bmatrix}
A_{T^c}^* ( A A^* )^{-1} \mathbb{Q} A_{T^c} x_{T^c}^{j} \\
= &
\begin{bmatrix}
(A_{T}^* A_{T})^{-1} A_{T}^* A_{T^c} A_{T^c}^* ( A A^* )^{-1} \mathbb{Q} \\
- A_{T^c}^* ( A A^* )^{-1} \mathbb{Q}
\end{bmatrix} A_{T^c} x_{T^c}^{j} .
\end{aligned}
\]
According to $A_{T^c} A_{T^c}^* = A A^* - A_T A_T^*$,
\[
\begin{aligned}
(A_{T}^* A_{T})^{-1} A_{T}^* A_{T^c} A_{T^c}^* ( A A^* )^{-1} \mathbb{Q}
= & (A_{T}^* A_{T})^{-1} A_{T}^* \left( A A^* - A_T A_T^* \right) ( A A^* )^{-1} \mathbb{Q} \\
= & (A_T^* A_T)^{-1} A_T^* \mathbb{Q} - A_T^* ( A A^* )^{-1} \mathbb{Q} \\
= & - A_T^* ( A A^* )^{-1} \mathbb{Q} .
\end{aligned}
\]
Therefore,
\[
\begin{aligned}
\Projection{\left( u^{j+1} - x^{j+1} \right)}
= &
\Projection{ \left(
\begin{bmatrix}
- A_T^* ( A A^* )^{-1} \mathbb{Q} \\
- A_{T^c}^* ( A A^* )^{-1} \mathbb{Q} \\
\end{bmatrix} A_{T^c} x_{T^c}^{j} \right) } \\
= & - \left( I - A^* (A A^*)^{-1} A \right) A^* ( A A^* )^{-1} \mathbb{Q} A_{T^c} x_{T^c}^{j} \\
= & 0.
\end{aligned}
\]
The claim is then followed.
\end{proof}

\begin{remark}
Theorem~\ref{T:NST_HT_FB_FixSupport} suggests that $T_{j+1} = T_{j}$ is a stopping criteria for NST+HT+FB. This theorem also explains partially why NST+HT+FB converges in finite steps.
\end{remark}

The next theorem touches on the convergence of the NST+HT algorithm, and establishes a convergence relationship between that of NST+HT+FB and that of NST+HT.

To avoid confusion in the next theorem, we denote by $\{ \tilde{T}_k \}$  the sequences of the index set corresponding to the $s$ most significant entries produced by the NST+HT+FB algorithm, and by $\{ \tilde{x}_k \}$ the corresponding solution.

\begin{theorem}\label{T:EquivalenceNST_HT_NST_HT_FB}
Let $A$ be a Parseval frame. Suppose $A_{T}^* A_{T}$ and $A_{T^c} A_{T^c}^*$ are invertible with $T \subset \{ 1, \ldots, N \}$ and $\Abs{T} = s$. Assume $T_i = T_{i+1} = \cdots = T$ for NST+HT for some integer $i$ and $\tilde{T}_j = \tilde{T}_{j+1}$ for NST+HT+FB for some integer $j$. If $x^{i} = \tilde{x}^{j}$, then $\lim_{k \to \infty} x^{k} = \tilde{x}^{j+1}$.
\end{theorem}

\begin{proof}
Since $x^{i} = \tilde{x}^{j}$, $T_i = \tilde{T}_j = T$. For NST+HT+FB, when $A$ is a Parseval frame, one can easily derive from Theorem~\ref{T:NST_HT_FB_FixSupport} that
\[
\tilde{x}^{j+1}_{T} = \tilde{x}^{j}_T + (A_T^* A_T)^{-1} A_T^* A_{T^c} \tilde{x}^{j}_{T^c}
\]
and
\[
\tilde{x}^{j+1}_{T^c} = A_{T^c}^* \left( I - A_T (A_T^* A_T)^{-1} A_T^* \right) A_{T^c} \tilde{x}^{j}_{T^c},
\]
which is exactly the limit of the sequence of $x^{k}$ in NST+HT, as demonstrated in Theorem~\ref{T:NST_HT_FixSupport}.
\end{proof}

Evidently, Theorem~\ref{T:EquivalenceNST_HT_NST_HT_FB} establishes the connections between NST+HT and NST+HT+FB. For Parseval frames $A$, the output of NST+HT+FB and the the limit of NST+HT are the same in this very sense.

In \cite{A_Candes_DecodingLinearProgramming} and various other articles, Cand\`{e}s and Tao introduced the notion of {\it restricted isometry property} (RIP) to analyze the performance of the solution to $(P_1)$. In the study of NST+HT and NST+HT+FB, it is natural to introduce a notion of {\it preconditioned restricted isometry property} (P-RIP) to analyze the performance of NST+HT and NST+HT+FB. To state our main results, we first recall the definition of restricted isometry constants.

\begin{definition}\label{D:RIP} \cite{A_Candes_DecodingLinearProgramming}
For each integers $s = 1,2,\ldots$, the restricted isometry constant $\delta_s$ of a matrix $A$ is defined as the smallest number $\delta_s$ such that
\begin{equation}\label{E:RIP}
(1-\delta_s) \Norm{x}{2}^2 \le \Norm{A x}{2}^2 \le (1+\delta_s) \Norm{x}{2}^2
\end{equation}
holds for all $s$-sparse vectors $x$.
\end{definition}

\begin{definition}\label{D:PreRIP}
For each integers $s = 1,2,\ldots$, the {\it preconditioned restricted isometry constant} $\gamma_s$ of a matrix $A$ is defined as the smallest number $\gamma_s$ such that
\begin{equation}\label{E:PRIP1}
(1-\gamma_s) \Norm{x}{2}^2 \le \Norm{(A A^*)^{-\frac{1}{2}} A x}{2}^2
\end{equation}
holds for all $s$-sparse vectors $x$.
\end{definition}

In fact, the preconditioned restricted isometry constant $\gamma_s$ characterizes the restricted isometry property of the preconditioned matrix $(A A^*)^{-\frac{1}{2}} A $. Since
\[
\Norm{(A A^*)^{-\frac{1}{2}} A x}{2} \le \Norm{(A A^*)^{-\frac{1}{2}} A}{2} \Norm{x}{2} = \Norm{x}{2},
\]$
\gamma_s$ is actually the smallest number such that, for all $s$-sparse vectors $x$,
\begin{equation}\label{E:PRIP2}
(1-\gamma_s) \Norm{x}{2}^2 \le \Norm{(A A^*)^{-\frac{1}{2}} A x}{2}^2 \le (1+\gamma_s) \Norm{x}{2}^2.
\end{equation}
That indicates $\gamma_s (A) = \delta_s ( (A A^*)^{-\frac{1}{2}} A )$. Evidently, for Parseval frames, since $A A^* = I$, $\gamma_s (A) = \delta_s (A)$. We shall term either \eqref{E:PRIP1} or \eqref{E:PRIP2} the {\it preconditioned restricted isometry property} (P-RIP).

\begin{remark}
We should note that the P-RIP is not particularly a stronger (than RIP) assumption over the matrix $A$.  In fact, an upper bound of the preconditioned restricted isometry constant $\gamma_s$ maybe expressed as a function of the RIP constant $\delta_s$ and the largest singular value, as stated in the following proposition.
\end{remark}

\begin{proposition}\label{T:UpperBoundPRIC}
Let $\sigma_{max}$ be the largest singular value of $A$. For each integers $s$, the P-RIP constant $\gamma_{s} \le 1 - \frac{1-\delta_s}{\sigma_{max}}$.
\end{proposition}

\begin{proof}
Since $\sigma_{max}$ is the largest singular value of $A$, $(A A^*)^{-1} - \frac{1}{\sigma_{max}} I$ is positive-semidefinite. Given any $s$-sparse vector $x$,
\[
\Norm{(A A^*)^{-\frac{1}{2}} A x}{2}^2 - \frac{1}{\sigma_{max}} \Norm{A x}{2}^2
= x^* A^* \left[ (A A^*)^{-1} - \frac{1}{\sigma_{max}} I \right] A x \ge 0.
\]
Consequently,
\[
\Norm{(A A^*)^{-\frac{1}{2}} A x}{2}^2 \ge \frac{1}{\sigma_{max}} \Norm{A x}{2}^2 \ge \frac{1 - \delta_s}{\sigma_{max}} \Norm{x}{2}^2,
\]
The upper bound of $\gamma_{s}$ is then followed.
\end{proof}

As a result, it is not hard to see from Proposition~\ref{T:UpperBoundPRIC} that the P-RIP is satisfied with high probability if $A$ is a (Gaussian) random matrix, just as in RIP.  We refer to studies of  extreme singular values of random matrices, e.g., \cite{A_Rudelson_ExtremeSingularValues, B_Vershynin_IntroductionNon-asymptoticAnalysisRandomMatrices}, for additional references.

\medskip

Before tuning to the convergence of the algorithms, let us also discuss a preferable expression of the P-RIP constants. Definition~\ref{D:PreRIP} actually implies $I - A_T^* (A A^*)^{-1} A_T \le \gamma_s I$, where $\Abs{T} \le s$. It then follows that $\Norm{I - A_T^* (A A^*)^{-1} A_T}{2} \le \gamma_{s}$, i.e., $\Norm{\mathbb{P}_{T,T}}{2} \le \gamma_{s}$. Here $\mathbb{P}_{T,T'}$ is the submatrix of the projection $\mathbb P$ consisting of rows indexed by $T$ and columns indexed by $T'$. Similarly, it follows from Definition~\ref{D:RIP} that $\Norm{I - A_T^* A_T}{2} \le \delta_{s}$.

Our main result is that, with requirements over the (preconditioned) restricted isometry constants of $A$, both procedures (NST+HT and NST+HT+FB) reduce the error in each iteration and are guaranteed to converge to limits with error bounds depending on the tail of the real solution and the noise.

\begin{theorem}\label{T:Convergence_NST_HT_Noisy}
Suppose $A x + e = b$ where $e$ is the measurement error or noise. Let $x^{\sharp}$ be the best $s$-term approximation of the real solution $x$. If the $(3 s)^{th}$ order P-RIP constant of $A$ satisfies $\gamma_{3s} < 0.5$, then $u^k$ in NST+HT satisfies
\[
\Norm{u^k - x^\sharp}{2} \le \rho^k \Norm{u^0 - x^\sharp}{2} + \frac{2}{1-\rho} \Norm{\tilde{e}}{2},
\]
where $\rho = 2 \gamma_{3s}$ and $\tilde{e} = (A A^*)^{-\frac{1}{2}} \left( A(x-x^\sharp)+e \right)$.
\end{theorem}

\begin{proof}
Let $T^\sharp$ be the index set corresponding to the $s$ most significant entries of the real solution, and let $T_k$ be the index set of the best $s$-term approximation $u^k$ of $x^k$.   Define $T \Define T^\sharp \cup T_k \cup T_{k-1}$. Obviously, $\Abs{T} \le 3s$.

For NST+HT, since $u^k$ is the best $s$-term approximation of $x^k$,
\[
\begin{aligned}
\Norm{u^k - x^k}{2} \le \Norm{x^\sharp - x^k}{2}
\Leftrightarrow
& \Norm{ (u^k - x^\sharp) + (x^\sharp - x^k) }{2}^2 \le \Norm{ x^\sharp - x^k }{2}^2 \\
\Leftrightarrow
& \Norm{u^k - x^\sharp}{2}^2 \le 2 \mathfrak{R} \InProd{u^k-x^\sharp}{x^k-x^\sharp} .
\end{aligned}
\]
With a slight abuse of notations, we then denote by $(u^{k-1}-x^\sharp)_T$ a vector consisting of the entries of $(u^{k-1}-x^\sharp)$ indexed by $T$ and zeros on $T^c$. Since $x^k = u^{k-1} + A^* (A A^*)^{-1} ( b - A u^{k-1} )$, as demonstrated in \eqref{E:EquivalentProjection}, and $A x^\sharp + A (x - x^\sharp) + e = b$, one has,
\[
\begin{aligned}
\Norm{u^k - x^\sharp}{2}^2
\le & 2 \mathfrak{R} \InProd{u^k-x^\sharp}{ u^{k-1} + A^* (A A^*)^{-1} ( b - A u^{k-1} ) - x^\sharp } \\
\le & 2 \mathfrak{R} \InProd{u^k-x^\sharp}{ \Projection{(u^{k-1}-x^\sharp)} + A^*(A A^*)^{-\frac{1}{2}} \tilde{e}} \\
= & 2 \mathfrak{R} \InProd{ (u^k-x^\sharp)_T }{ [ \Projection{ (u^{k-1}-x^\sharp)_T } ]_T } +  2 \mathfrak{R} \InProd{u^k-x^\sharp}{A^*(A A^*)^{-\frac{1}{2}} \tilde{e}} \\
\le & 2 \Norm{u^k-x^\sharp}{2} \Norm{\mathbb P_{T,T}}{2} \Norm{u^{k-1}-x^\sharp}{2} + 2 \Norm{u^k-x^\sharp}{2} \Norm{\tilde{e}}{2},
\end{aligned}
\]
where $\mathfrak R$ stands for the operation of taking the real part of a variable.  It then follows that
\[
\begin{aligned}
\Norm{u^k - x^\sharp}{2}
\le & 2 \Norm{\mathbb P_{T,T}}{2} \Norm{u^{k-1}-x^\sharp}{2} + 2 \Norm{\tilde{e}}{2} \\
\le & 2 \gamma_{3s} \Norm{u^{k-1}-x^\sharp}{2} + 2 \Norm{\tilde{e}}{2}\\
= & \rho \Norm{u^{k-1}-x^\sharp}{2} + 2 \Norm{\tilde{e}}{2} .
\end{aligned}
\]
Therefore,
\[
\Norm{u^k - x^\sharp}{2} \le \rho^k \Norm{u^{0}-x^\sharp}{2} + \frac{2}{1-\rho} \Norm{\tilde{e}}{2}.
\]
The claim is then followed.
\end{proof}

One can easily reach the following corollary characterizing the behavior of NST+HT in the noiseless case where $x$ is exactly sparse.

\begin{corollary}\label{T:Convergence_NST_HT}
Let $x$ be the solution to $A x = b$ with only $s$ sparsity. If the $(3s)^{th}$ order P-RIP constant of $A$ satisfies $\gamma_{3s} < 0.5$, then the sequence of $u^k$ in NST+HT converges to $x$.
\end{corollary}

\begin{theorem}\label{T:Convergence_NST_HT_FB_Noisy}
Suppose $A x + e = b$ with the measurement error $e$.  Let $x^{\sharp}$ be the best $s$-term approximation of the real solution $x$. If the P-RIP and RIP constants of $A$ satisfy $\delta_{2s} + \sqrt{2} \gamma_{3s} < 1$, then $u^k$ in NST+HT+FB satisfies
\[
\Norm{u^k - x^\sharp}{2} \le \rho^k \Norm{u^0 - x^\sharp}{2} + \frac{\tau}{1-\rho} \Norm{\tilde{e}}{2},
\]
where $\rho = \frac{ \sqrt{2} \gamma_{3s} }{ 1-\delta_{2s} }$, $\tau = \frac{ \sqrt{2} + \sqrt{1+\delta_s} }{1-\delta_{2s}}$ and $\tilde{e} = A (x-x^\sharp) + e$.
\end{theorem}

\begin{proof}
Similar to the proof of Theorem~\ref{T:Convergence_NST_HT_Noisy}, we denote by $T^\sharp$ the index set corresponding to the most $s$ significant entries of the real solution and let $x_T$ be the vector which keeps only the entries indexed by $T$. For convenience, we define $T \Define T^\sharp \cup T_k$ and $\mathbb Q \Define A^* (A A^*)^{-\frac{1}{2}} A$, respectively. Evidently, $\Abs{T} \le 2s$ and $\Projection{} = I - \mathbb Q$.

According to the iteration of NST+HT+FB, for any $z \in \R^N$ being supported on $T_k$,
\[
\begin{aligned}
\InProd{A u^k - b}{A z}
= & \InProd{A_{T_k} x^k_{T_k} + A _{T_k} (A_{T_k}^* A_{T_k})^{-1} A_{T_k}^* A_{T_k^c} x^k_{T_k^c} - b}{A_{T_k} z_{T_k}} \\
= & \InProd{A_{T_k}^* (A_{T_k} x^k_{T_k} + A_{T_k^c} x^k_{T_k^c} - b) }{z_{T_k}}
= 0 .
\end{aligned}
\]
The last step is due to the feasibility of $x^k$. The inner product can also be written as
\[
\begin{aligned}
\InProd{A u^k - b}{A z}
= & \InProd{A u^k - A x^\sharp - A (x - x^\sharp) - e}{A z} \\
= & \InProd{A (u^k - x^\sharp) - \tilde{e}}{A z}
= 0 .
\end{aligned}
\]
Therefore, $\InProd{u^k - x^\sharp}{A^* A z} = \InProd{\tilde{e}}{A z}$.

With a slight abuse of notations, we denote by $(u^k - x^\sharp)_{T_k}$ a vector consisting of the entries of $(u^k - x^\sharp)$ indexed by $T_k$ and zeros on $T_k^c$. Clearly, $(u^k - x^\sharp)_{T_k}$ is supported on $T_k$. Therefore,
\[
\InProd{u^k - x^\sharp}{A^* A (u^k - x^\sharp)_{T_k}} = \InProd{ \tilde{e} }{A (u^k - x^\sharp)_{T_k}}.
\]
Consequently,
\[
\begin{aligned}
\Norm{(u^k - x^\sharp)_{T_k}}{2}^2
= & \InProd{u^k - x^\sharp}{(u^k - x^\sharp)_{T_k}} \\
= & \InProd{u^k - x^\sharp}{(I - A^* A)(u^k - x^\sharp)_{T_k}} + \InProd{\tilde{e}}{A (u^k - x^\sharp)_{T_k})} \\
\le & \Norm{u^k - x^\sharp}{2} \Norm{I - A_T^* A_T}{2} \Norm{(u^k - x^\sharp)_{T_k}}{2} + \Norm{\tilde{e}}{2} \Norm{A (u^k - x^\sharp)_{T_k}}{2} \\
\le & \delta_{2s} \Norm{u^k - x^\sharp}{2} \Norm{(u^k - x^\sharp)_{T_k}}{2} + \sqrt{1+\delta_s} \Norm{\tilde{e}}{2} \Norm{(u^k - x^\sharp)_{T_k}}{2} .
\end{aligned}
\]
It then follows that
\[
\Norm{(u^k - x^\sharp)_{T_k}}{2} \le \delta_{2s} \Norm{u^k - x^\sharp}{2} + \sqrt{1+\delta_s} \Norm{\tilde{e}}{2} .
\]
Since $u^k$ and $x^\sharp$ are supported on $T^k$ and $T^\sharp$ respectively,
\[
\begin{aligned}
\Norm{u^k - x^\sharp}{2}
= & \Norm{(u^k - x^\sharp)_{T_k} + (u^k - x^\sharp)_{T \backslash T_k}}{2} \\
\le & \Norm{(u^k - x^\sharp)_{T_k}}{2} + \Norm{ u^k_{T \backslash T_k} - x^\sharp_{T \backslash T_k} }{2} \\
\le & \delta_{2s} \Norm{u^k - x^\sharp}{2} + \sqrt{1+\delta_s} \Norm{\tilde{e}}{2} + \Norm{ x^\sharp_{T \backslash T_k}}{2} .
\end{aligned}
\]
Therefore,
\[
\Norm{u^k - x^\sharp}{2}^2 \le \frac{1}{1-\delta_{2s}} \Norm{x^\sharp_{T \backslash T_k}}{2} + \frac{\sqrt{1+\delta_s}}{1-\delta_{2s}} \Norm{\tilde{e}}{2} .
\]

Recall that $T_k$ corresponds to the most $s$ significant entries of $x^k$, it follows that
\[
\Norm{x^k_{T^\sharp}}{2}^2 \le \Norm{x^k_{T^k}}{2}^2.
\]
Exploiting the fact $x^{k} = x^{k-1} + \Projection{(u^{k-1} - x^{k-1})}$, and eliminating the common terms over $T^\sharp \cap T^k$, one has
\begin{equation}\label{E:NST_HT_FB_1}
\Norm{ [ x^{k-1} + \Projection{(u^{k-1} - x^{k-1})} ]_{T^\sharp \backslash T_k}}{2} \le \Norm{ [ x^{k-1} + \Projection{(u^{k-1} - x^{k-1})} ]_{T_k \backslash T^\sharp}}{2} .
\end{equation}
Since $A x^{k-1} = A x^\sharp + \tilde{e} = b$, for the left side of \eqref{E:NST_HT_FB_1},
\[
\begin{aligned}
& \Norm{ [ x^{k-1} + \Projection{(u^{k-1} - x^{k-1})} ]_{T^\sharp \backslash T_k}}{2} \\
= & \Norm{ [ u^{k-1} + A^* (A A^*)^{-\frac{1}{2}} ( A x^{k-1} - A u^{k-1} ) ]_{T^\sharp \backslash T_k}}{2} \\
= & \Norm{ [ u^{k-1} + A^* (A A^*)^{-\frac{1}{2}} ( A x^\sharp + \tilde{e} - A u^{k-1} ) ]_{T^\sharp \backslash T_k}}{2} \\
= & \Norm{ [ u^{k-1} + \mathbb Q (x^\sharp - u^{k-1}) ]_{T^\sharp \backslash T_k} + [ A^* (A A^*)^{-\frac{1}{2}} \tilde{e} ]_{T^\sharp \backslash T_k} }{2} .
\end{aligned}
\]
For the right side of \eqref{E:NST_HT_FB_1},
\[
\begin{aligned}
& \Norm{ [ x^{k-1} + \Projection{(u^{k-1} - x^{k-1})} ]_{T_k \backslash T^\sharp}}{2} \\
= & \Norm{ [ u^{k-1} + A^* (A A^*)^{-\frac{1}{2}} (A x^{k-1} - A u^{k-1}) ]_{T_k \backslash T^\sharp}}{2} \\
= & \Norm{ [ u^{k-1} + A^* (A A^*)^{-\frac{1}{2}} (A x^\sharp - A u^{k-1} + \tilde{e}) ]_{T_k \backslash T^\sharp}}{2} \\
= & \Norm{ [ \Projection{(u^{k-1} - x^\sharp)} ]_{T_k \backslash T^\sharp} + [ A^* (A A^*)^{-\frac{1}{2}} \tilde{e} ]_{T_k \backslash T^\sharp} }{2}.
\end{aligned}
\]
Then \eqref{E:NST_HT_FB_1} is equivalent to
\[
\begin{aligned}
& \Norm{ [ u^{k-1} + \mathbb Q (x^\sharp - u^{k-1}) ]_{T^\sharp \backslash T_k} + [ A^* (A A^*)^{-\frac{1}{2}} \tilde{e} ]_{T^\sharp \backslash T_k} }{2} \\
\le & \Norm{ [ \Projection{(u^{k-1} - x^\sharp)} ]_{T_k \backslash T^\sharp} + [ A^* (A A^*)^{-\frac{1}{2}} \tilde{e} ]_{T_k \backslash T^\sharp} }{2}
\end{aligned}
\]
Furthermore, by the triangle inequality,
\[
\begin{aligned}
\Norm{ [ u^{k-1} + \mathbb Q (x^\sharp - u^{k-1}) ]_{T^\sharp \backslash T_k}}{2}
\le & \Norm{ [ \Projection{(u^{k-1} - x^\sharp)} ]_{T_k \backslash T^\sharp}}{2} + \Norm{ [ A^* (A A^*)^{-\frac{1}{2}} \tilde{e} ]_{T_k \backslash T^\sharp} }{2} \\
& + \Norm{ [ A^* (A A^*)^{-\frac{1}{2}} \tilde{e} ]_{T^\sharp \backslash T_k} }{2}.
\end{aligned}
\]
Together with the decomposition
\[
x^\sharp_{T^\sharp \backslash T_k} = [ x^\sharp - u^{k-1} - \mathbb Q (x^\sharp - u^{k-1}) ]_{T^\sharp \backslash T_k} + [ u^{k-1} + \mathbb Q (x^\sharp - u^{k-1}) ]_{T^\sharp \backslash T_k},
\]
we have
\[
\begin{aligned}
\Norm{ x^\sharp_{T^\sharp \backslash T_k} }{2}
\le & \Norm{ [ x^\sharp - u^{k-1} - \mathbb Q (x^\sharp - u^{k-1}) ]_{T^\sharp \backslash T_k} }{2} + \Norm{ [ u^{k-1} + \mathbb Q (x^\sharp - u^{k-1}) ]_{T^\sharp \backslash T_k} }{2}\\
\le & \Norm{ [ \Projection{(x^\sharp - u^{k-1})} ]_{T^\sharp \backslash T_k} }{2} + \Norm{ [ \Projection{(u^{k-1} - x^\sharp)} ]_{T_k \backslash T^\sharp}}{2} \\
& + \Norm{ [ A^* (A A^*)^{-\frac{1}{2}} \tilde{e} ]_{T_k \backslash T^\sharp} }{2} + \Norm{ [ A^* (A A^*)^{-\frac{1}{2}} \tilde{e} ]_{T^\sharp \backslash T_k} }{2} \\
\le & \sqrt{2} \left( \Norm{[ \Projection{ (x^\sharp - u^{k-1}) } ]_{T}} {2} + \Norm{[ A^* (A A^*)^{-\frac{1}{2}} \tilde{e} ]_{T} }{2} \right) \\
\le & \sqrt{2} \left( \Norm{ \mathbb P _{T \cup T_{k-1},T \cup T_{k-1}} }{2} \Norm{x^\sharp - u^{k-1}}{2} + \Norm{A^* (A A^*)^{-\frac{1}{2}}}{2} \Norm{\tilde{e}}{2} \right) \\
\le & \sqrt{2} \left( \gamma_{3s} \Norm{x^\sharp - u^{k-1}}{2} + \Norm{\tilde{e}}{2} \right) .
\end{aligned}
\]
Therefore,
\[
\Norm{u^k - x^\sharp}{2}^2 \le \frac{ \sqrt{2} \gamma_{3s} }{ 1-\delta_{2s} } \Norm{u^{k-1} - x^\sharp}{2} + \frac{ \sqrt{2} + \sqrt{1+\delta_s} }{1-\delta_{2s}} \Norm{\tilde{e}}{2} .
\]
It then follows that
\[
\Norm{u^k - x^\sharp}{2}^2 \le \rho^k \Norm{u^0 - x^\sharp}{2} + \frac{\tau}{1-\rho} \Norm{\tilde{e}}{2} ,
\]
where $\rho = \frac{ \sqrt{2} \gamma_{3s} }{ 1-\delta_{2s} }$ and $\tau = \frac{ \sqrt{2} + \sqrt{1+\delta_s} }{1-\delta_{2s}}$.
\end{proof}

In the previous proof, we have made use of some techniques introduced by Foucart in \cite{A_Foucart_HardThresholdingPursuit, A_Foucart_SparseRecoveryAlgorithms}. It is also possible to improve or refine the condition $\delta_{2s} + \sqrt{2} \gamma_{3s} < 1$, as demonstrated in \cite{A_Foucart_HardThresholdingPursuit}, though it is not a focus of this article.

Let us also observe that if $x$ is exactly sparse without measurement noise, then the NST+HT+FB algorithm converges in finite steps, which really places this algorithm among the very fast algorithms known up to date.

\begin{theorem}\label{T:Convergence_NST_HT_FB}
Let $x$ be the solution to $Ax=b$ with $s$ sparsity. If the P-RIP and RIP constants of $A$ satisfies $\delta_{2s} + \sqrt{2} \gamma_{3s} < 1$, then the sequence of $u^k$ in NST+HT+FB converges to $x$ in finite many steps.
\end{theorem}

\begin{proof}
Since $u^{k}$ consists of only $s$ non-zero entries and the sequence of $u^{k}$ converges to $x$, there must exist a sufficiently large integer $j$ such that $T_{j}$ is the support of $x$. Then, by Theorem~\ref{T:NST_HT_FB_FixSupport}, the immediate next step of NST+HT+FB gives rise to $u^{j+1} = x$. The claim is followed.
\end{proof}

\begin{remark}
%
%

Although NST+HT+FB has a pursuit spirit seen in various algorithms such as HTP \cite{A_Foucart_HardThresholdingPursuit}, we should not undervalue the feedback form that we proposed here.  The feedback mechanism plays a significant role particularly for large scale problems.

In fact, the point of feedback enables us to derive several other more efficient algorithms, as presented in Section~\ref{S:OtherChoices}.  For large scale problems, these suboptimal algorithms based on the feedback idea, NST+HT+subFB and NST+stretchedHT, are seen much more realistic and much faster without having to compute inverses $(A_{T_k}^* A_{T_k})^{-1}$ in each and every step.
\end{remark}

\section{Effectiveness at Large Scales and Numerical examples}\label{S:NumericalExamples}
In this section, we support the claim that the proposed algorithms are very effective (particularly at large scales) in practice by extensive numerical experiments in four different aspects.
\begin{itemize}
\item The first is to illustrate the overall performance of the NST based algorithms in terms of execution-time and comparisons with known fast algorithms.
\item The second is to investigate the frequency of exact/successful recovery of the NST based algorithms and comparison with known effective algorithms.
\item The third is about the performance comparison of the algorithms in the noisy cases among the NST based algorithms and others.
\item The fourth is to demonstrate the performance differentiations within the class of the NST based algorithms.
\end{itemize}

Among many, by known algorithms we mean those that are similar to the NST based approaches.  These include IHT, OMP, HTP and SP, which are all methods attempting to address the sparse solutions. One BP algorithm, CVX \cite{CVX}, is also included in the comparison to understand the solution capacity between the $\ell_0$-minimization and the $\ell_1$-minimization methodologies.  The comparisons with these representative algorithms are to demonstrate the performance differentiations, and to provide a measurement scale and judgement foundation for the algorithms comparison.

We must comment that the comparisons are far from complete. There are indeed some algorithms that we decide not to involve in the comparison. These include, for instance, CoSaMP and linearized Bregman iteration. For CoSaMP, the reason is that the main idea and the performance of CoSaMP and SP are very similar. For the Bregman iteration, the reasons are twofold.  One is that Bregman iterations are a means to find an approximate solution to $(P_1)$, while our algorithms attempt to solve an $\ell_0$-minimization problem. Solutions to $(P_1)$ is represented by the CVX in our studies (mostly for the existence of the solution, not for the speed of the algorithm). The other is due to the fact that the optimal performances of various Bregman iterations require somewhat skillful selections of the parameters. It is not easy to ensure a fair comparison. However, from the approximate spirit and the algorithm formulation of the Bregman iterations, we suspect that the NST based algorithms outperform the linearized Bregman iteration substantially. Our many trials of a known linearized Bregman algorithm with kicking confirm such believes. We did not include these trials in the article because we were not entirely sure if the parameters of the Bregman iterations have been selected optimally.

Except for the specific large scale tests, most tests uses a $128 \times 256$ matrix $A$ with standard i.i.d. Gaussian entries. All the columns are normalized to unit $\ell_2$-norm. The support of the sparse signal is chosen randomly. We refer Gaussian sparse signals to those whose nonzero entries are drawn independently from the Gaussian distribution with zero mean and unit variance. For Bernoulli sparse signals, the nonzero entries are drawn independently from $\pm 1$ with equiprobability.

\subsection{Overall execution-time comparison}
\begin{figure}
\centering
\subfigure[Small scale problems, $n=128$ and $N=256$.]{
\label{F:CPUTimeGaussianGaussian:SmallScale}
\includegraphics[width=0.485\columnwidth]{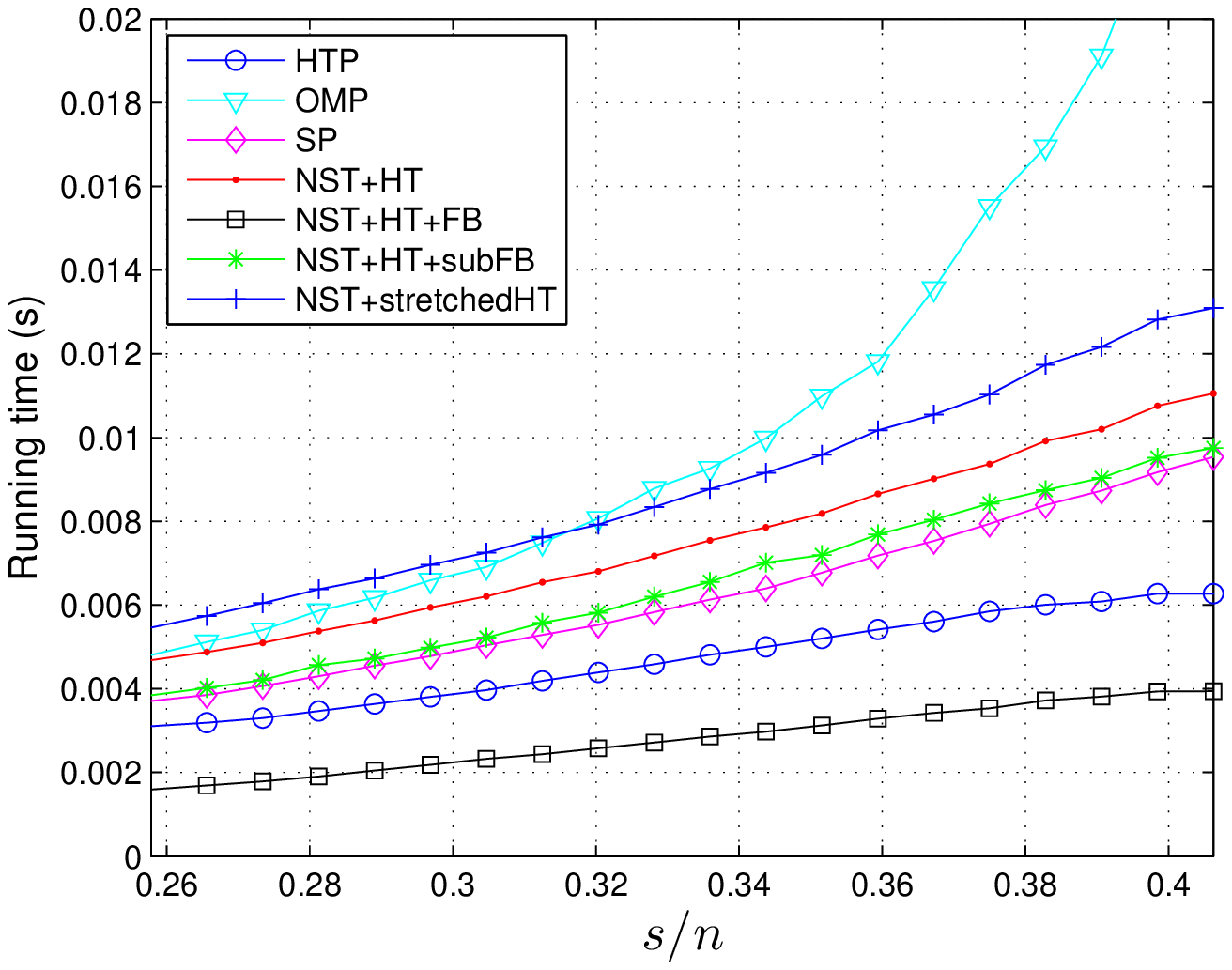}}
\subfigure[Large scale problems, $n=5000$ and $N=10000$.]{
\label{F:CPUTimeGaussianGaussian:LargeScale}
\includegraphics[width=0.485\columnwidth]{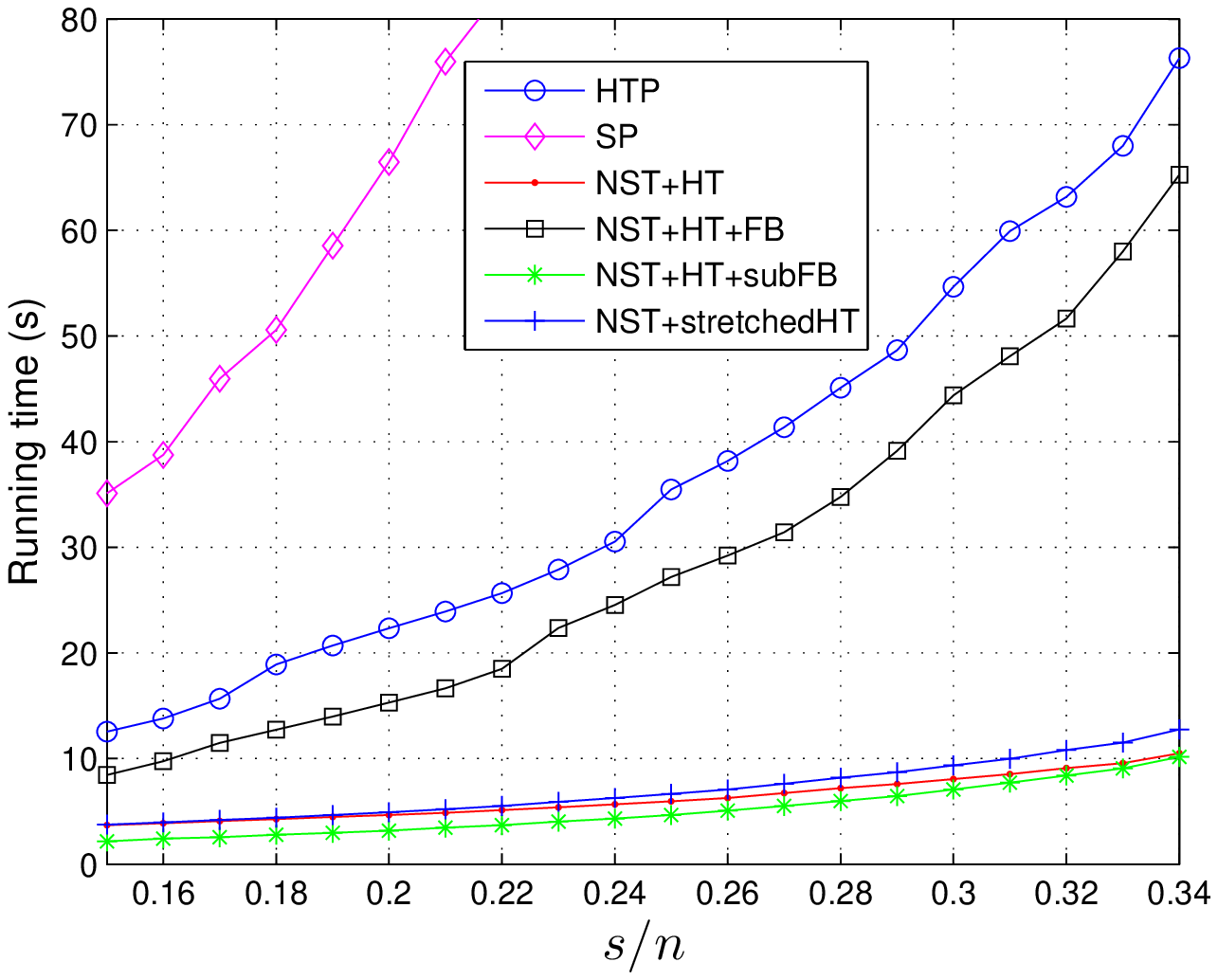}}
\caption{Plots of the execution-time as a function of $s/n$. (a) For small scale problems, NST+HT+FB is the most efficient. (b) For large scale problems, the NST based algorithms (not including NST+HT+FB) are more than 5 times faster than the other algorithms.}
\label{F:CPUTimeGaussianGaussian}
\end{figure}

For execution-time comparison, we tested both the small scale ($128\times 256$) and the large scale ($5000\times 10000$) problems.  For each sparsity value $s$, the small scale is tested for 5000 trials, while the large scale is tested for 20 trials. The execution-time is recorded for every trial. The average time is then calculated and plotted in Figure~\ref{F:CPUTimeGaussianGaussian}. Note, we have set $\epsilon_1 = 10^{-5}$ and $\epsilon_2 = 10^{-6}$ as stopping parameters for all the NST based algorithms. The parameter $\lambda$ in the NST+HT+subFB algorithm has been set to $\lambda=1$ in all experiments.

For small scale problems, Figure~\ref{F:CPUTimeGaussianGaussian:SmallScale} indicates that NST+HT+FB is the most efficient among all the methods compared.  The HTP \cite{A_Foucart_HardThresholdingPursuit} stands at the second position. The other three NST based algorithms spend a little more time than SP. The execution-time of OMP grows dramatically as the sparsity (number of nonzeros) increases.

For large scale problems, however, Figure~\ref{F:CPUTimeGaussianGaussian:LargeScale} shows that the NST based algorithms NST+HT, NST+HT+subFB, and NST+stretchedHT are much more efficient (more than 5 times faster) than all competitive algorithms.  The efficiency is particularly evident as the sparsity level increases.  The execution-time of the three NST based algorithms grows much slower than those of other algorithms as the sparsity value increases.

\subsection{Overall successful recovery performance and comparison}
The second test is to compare the frequency of exact recovery of the NST based algorithms with those of other algorithms. For the parameters of the adaptive NST based algorithms, we refer to the last test set. For Gaussian sparse vectors, we set the initial sparsity value $s_0=0.5s$ for the adaptive NST+stretchedHT algorithm. For the other three adaptive NST based algorithms, we set $s_0=0.3s$. For Bernoulli sparse vectors, we set $s_0 = 0.9s$ for all the adaptive algorithms. To obtain better performance, we set $s'=1$ for all scenarios. Each algorithm is tested for $500$ trials for every value of $s$. An exact recovery is then recorded whenever $\Norm{u - x}{2}/\Norm{x}{2} \le 10^{-4}$. The frequency of the exact recovery as a function of the sparsity measurement ratio $s/n$ is plotted.

\begin{figure}
\centering
\subfigure[Gaussian sparse vectors.]{
\label{F:RecoveryFrequencyGaussianGaussian}
\includegraphics[width=0.485\columnwidth]{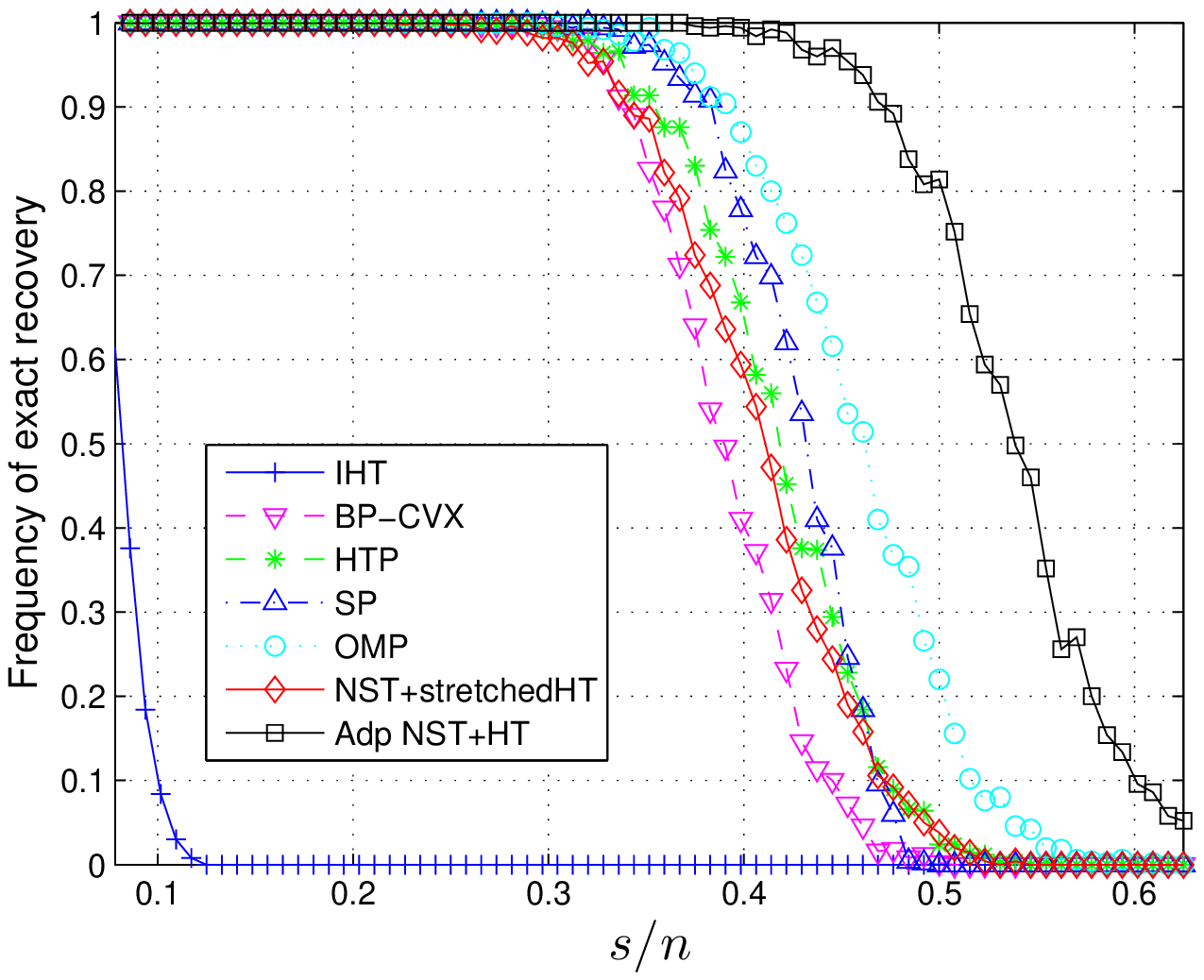}}
\subfigure[Bernoulli sparse vectors.]{
\label{F:RecoveryFrequencyGaussianSign}
\includegraphics[width=0.485\columnwidth]{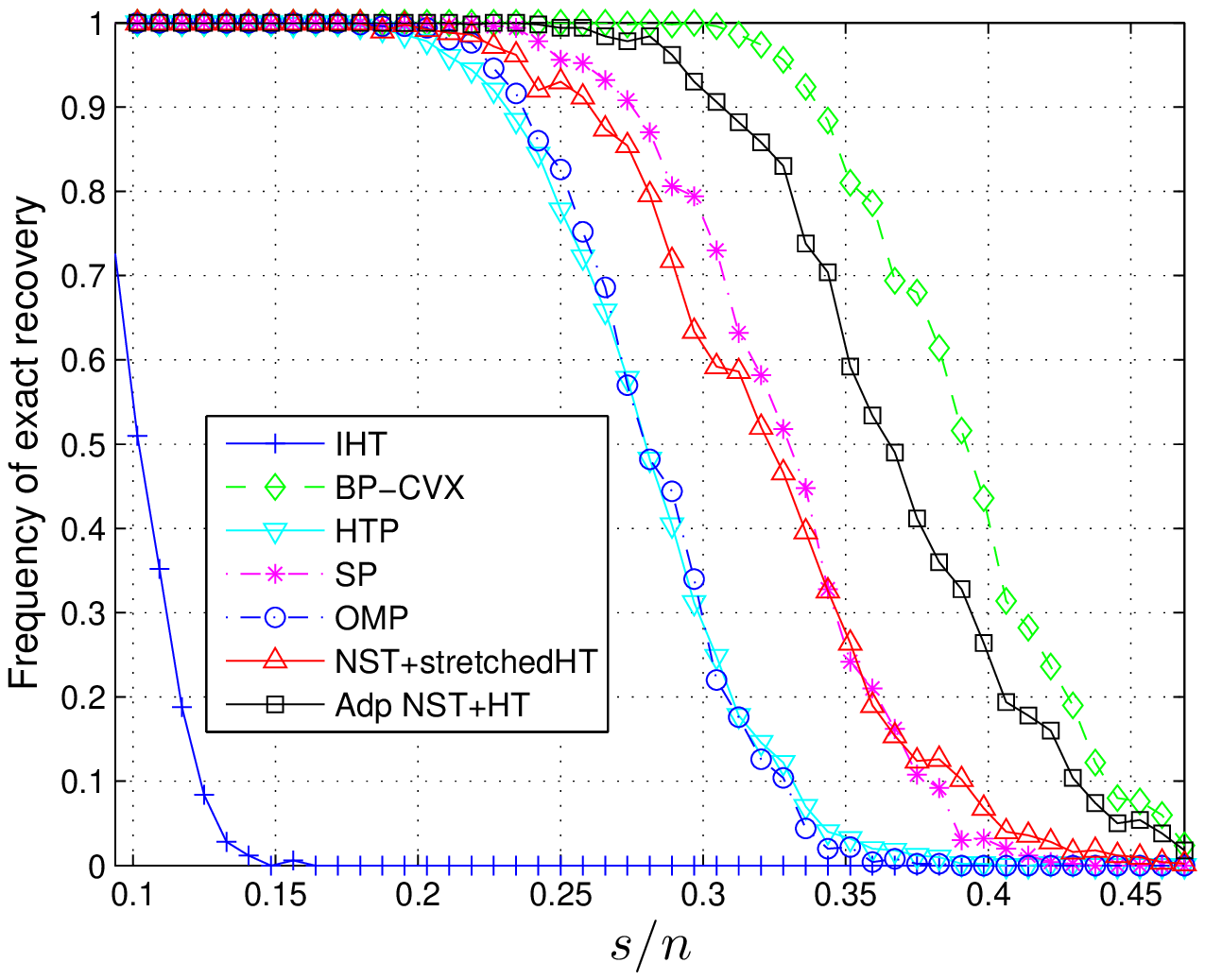}}
\caption{(a) Plots of the frequency of exact recovery of \emph{Gaussian sparse vectors} as a function of $s/n$. Exact recovery means $\Norm{u - x}{2}/\Norm{x}{2} \le 10^{-4}$. The adaptive NST based algorithms outperform other algorithms greatly. (b) Plots of the frequency of exact recovery of \emph{Bernoulli sparse vectors} as a function of $s/n$. BP-CVX possesses the best performance and the adaptive NST based algorithms become competitive.}
\label{F:RecoveryFrequency}
\end{figure}

The results for Gaussian sparse signals are demonstrated in Figures~\ref{F:RecoveryFrequencyGaussianGaussian}. It is evident that the adaptive NST based algorithms outperform all other algorithms by a great margin. The performance of the nonadaptive NST based algorithms is similar to that of HTP. For the clarity of the plots, the results of the other three (adaptive) NST based are not plotted here. Their performances are illustrated in Figure~\ref{F:RecoveryFrequencyNST_GaussianGaussian} in the last numerical experiment.

Unlike the situation of recovering Gaussian sparse vectors, for Bernoulli sparse vectors, BP-CVX performs the best, as demonstrated in Figure~\ref{F:RecoveryFrequencyGaussianSign}. Such an advantage of BP for recovering Bernoulli sparse signals is also observed in \cite{A_Dai_SubspacePursuit, A_Foucart_HardThresholdingPursuit}.

The adaptive NST+HT algorithm is the next best algorithm, still outperforms OMP, HTP and SP significantly.  It is also clear that the nonadaptive NST based algorithms outperform OMP and HTP by a good margin as well. Evidently, all the algorithms outperform IHT greatly in both scenarios,

\subsection{Comparison in contaminated signals and noisy measurements}
This study is to investigate the performance of the NST based and other algorithms in the noisy cases, which consists of two numerical experiments. The first is to deal with the contaminated signals case. To guarantee fixed signal-to-noise ratio (SNR), the ideal sparse signal $x^\sharp$, which is contaminated with zero-mean white Gaussian noise $v$, is normalized to unit $\ell_2$-norm. We rescale the noise level $\varepsilon \Define \Norm{v}{2}$ to the specified values and then obtain $b = A (x^\sharp + v)$. The second experiment is to deal with the contaminated measurement case. The ideal measurement $A x^\sharp$ is normalized to unit $\ell_2$-norm and contaminated with zero-mean white Gaussian noise $v$, whose level $\varepsilon \Define \Norm{v}{2}$ is rescaled to the specified values. The measurement is then $b = A x^\sharp + v$. In all the three experiments, $5000$ trials are performed with the noise level $\varepsilon = 0, 0.01, \ldots, 0.2$. To guarantee the best performance of OMP \cite{A_Tropp_GreedIsGood}, only $s$ iterations are carried out for every trial.

\begin{figure}
\centering
\subfigure[Signal is contaminated: $A(x^{\sharp}+v) = b$]{
\label{F:RelativeErrorNoisy:Signal}
\includegraphics[width=0.485\columnwidth]{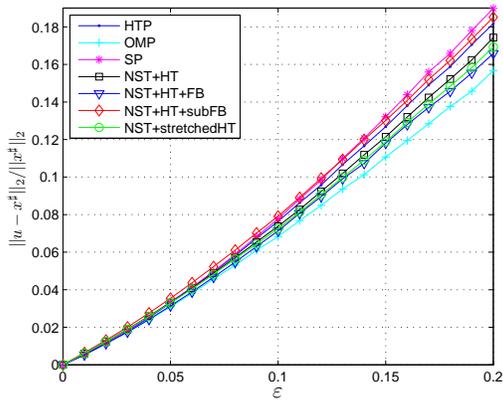}}
\subfigure[Measurement is contaminated: $Ax^{\sharp}+v = b$]{
\label{F:RelativeErrorNoisy:Measurement}
\includegraphics[width=0.485\columnwidth]{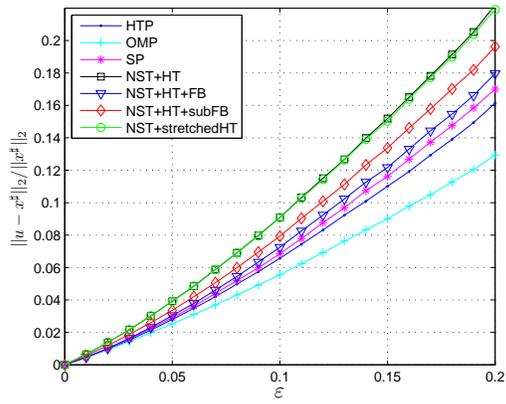}}
\caption{Plots of $\Norm{u - x^\sharp}{2}/\Norm{x^\sharp}{2}$ as a function of the noise level $\varepsilon$ for all the NST based algorithms and other algorithms. Here $s = 20$. OMP outperforms the other algorithms in both cases, but the differentiation is not dramatic when the signals are contaminated.}
\label{F:RelativeErrorNoisy}
\end{figure}

The results are demonstrated in Figure~\ref{F:RelativeErrorNoisy}, which implies that all algorithms possess better stabilities. It's clearly that OMP outperforms the other algorithms in both cases, but the differentiation is not dramatic when the signals are contaminated. In the NST group, for both contaminated signals and noisy measurements, NST+HT+FB outperforms the other three algorithms slightly.

\subsection{Performance comparison within the class of the NST based algorithms}
We test the overall performance of the NST based algorithms in three numerical experiments. The first is to demonstrate the convergence. With fixed value of $s$, for each of $5000$ trials, we record the relative error $\Norm{u^{k} - x}{2} / \Norm{x}{2}$ for every $k$.

\begin{figure}
\centering
\subfigure[The relative error curves of $10$ instances of NST+HT+FB.]{
\label{F:RelativeErrorNST:10Instances}
\includegraphics[width=0.485\columnwidth]{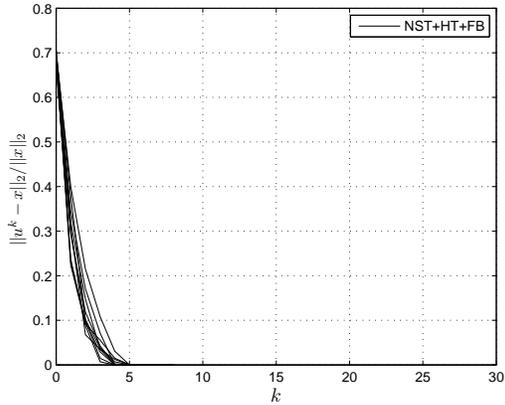}}
\subfigure[The average relative error curves.]{
\label{F:RelativeErrorNST:GaussianGuassian}
\includegraphics[width=0.485\columnwidth]{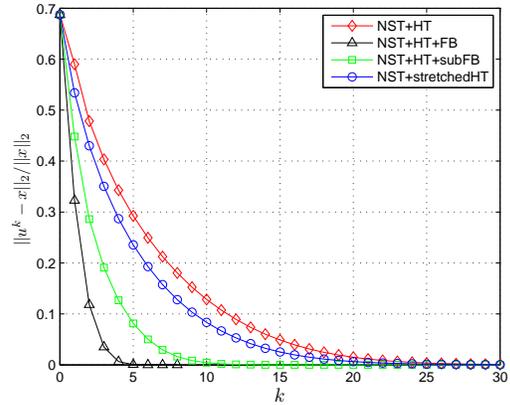}}
\caption{Plots of the error $\Norm{u^{k} - x}{2} / \Norm{x}{2}$ for recovering \emph{Gaussian sparse vectors} as a function of $k$ for all the NST based algorithms. Here $s=30$. It takes NST+HT+FB no more than $10$ steps for exact recovery. The initial relative errors are all set as $\Norm{x^0-x}{2}/\Norm{x}{2}$.}
\label{F:RelativeErrorNST}
\end{figure}

Figure~\ref{F:RelativeErrorNST:10Instances} shows the fact that the NST+HT+FB algorithm converges in finite many steps.  It demonstrates the first 10 instances of the 5000 trials for recovering Gaussian sparse vectors. Figure~\ref{F:RelativeErrorNST:GaussianGuassian} plots the average errors of all the NST based algorithms as functions of $k$.

\begin{figure}
\centering
\includegraphics[width=0.5\columnwidth]{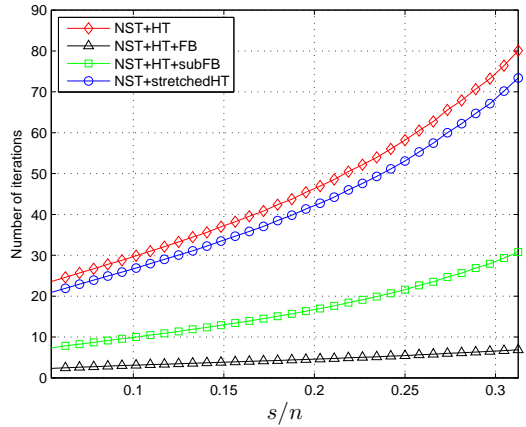}
\caption{Plots of the number of iterations as a function of $s/n$. Clearly, it takes NST+HT+FB far less number of iterations for convergence.}
\label{F:NumberIterations}
\end{figure}

The second experiment is to illustrate the number of iterations for convergence. For each value of $s$, we first recorded the number for all of $5000$ trials. The average number is then calculated. Figure~\ref{F:NumberIterations} plots the average number as a function of the sparsity measurement ratio $s/n$.

The obvious conclusion from the first two experiments is that NST+HT+FB converges dramatically rapidly than the other three NST based algorithms for small scale problems.  We are surprised that, in this setting ($s = 30$, $s/n \approx 0.23$), it always takes NST+HT+FB no more than $10$ steps to recover sparse vectors exactly. In addition, NST+HT+subFB and NST+stretchedHT all possess better convergent behaviors than NST+HT. The first two experiments verify the roles of the feedback and the stretch, evidently.

The third experiment is to examine the performance of the adaptive NST based algorithms with different initial sparsity values $s_0$ in the noiseless case. For simplicity, we just set $s_0 = \kappa s$ and $s'=1$. Take adaptive NST+HT as examples. With fixed $s$ and $0 < \kappa < 1$, we perform adaptive NST+HT $500$ trials. An exact recovery is then recorded whenever $\Norm{u - x}{2}/\Norm{x}{2} \le 10^{-4}$. Finally, we plot the frequency of exact recovery as a function of $s/n$ for different values of $\kappa$.

\begin{figure}
\centering
\subfigure[Gaussian sparse vectors]{
\label{F:RecoveryFrequencyAdaptive:GaussianVectors}
\includegraphics[width=0.485\columnwidth]{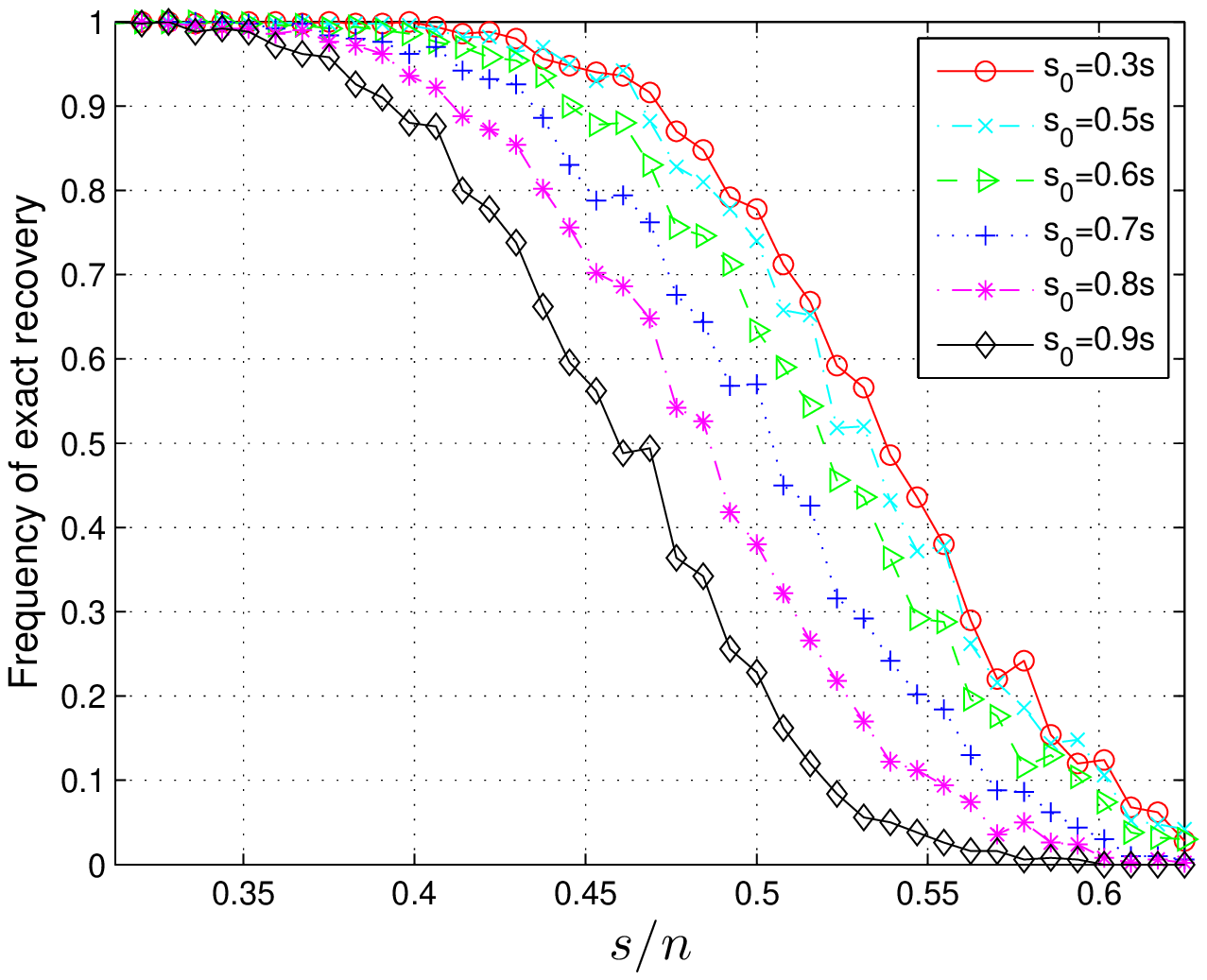}}
\subfigure[Bernoulli sparse vectors]{
\label{F:InitializationAdaptive:BernoulliVectors}
\includegraphics[width=0.485\columnwidth]{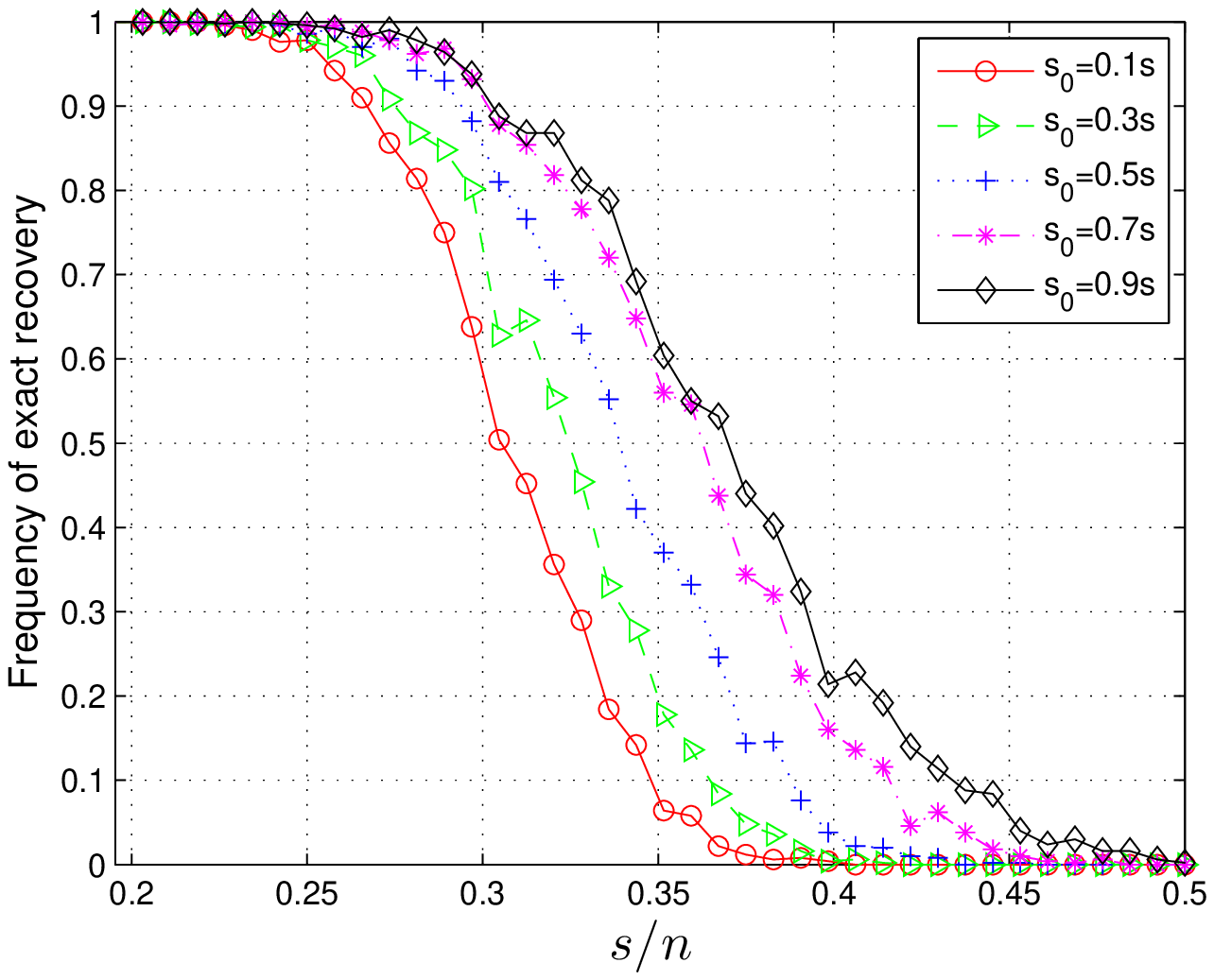}}
\caption{Investigation of the initial sparsity values: plots of the frequency of exact recovery of sparse vectors as a function of $s/n$ for the adaptive NST+HT algorithm. (a) Recover \emph{Gaussian sparse vectors}. The value $s_0=0.3s$ outperforms the other choices.  Smaller than $0.3s$ value of $s_0$ would not make much difference. (b) Recover \emph{Bernoulli sparse vectors}.  The value $s_0=0.9s$ outperforms the other choices.}
\label{F:InitializationAdaptive}
\end{figure}

Figure~\ref{F:InitializationAdaptive} demonstrates the frequency of exact recovery as a function of $s/n$ for different $\kappa$. We actually recorded the results for $\kappa=0.1,0.2,\ldots,0.9$. However, for better contrast of the plots, the representative curves within the transition range are only presented.

The observation is that the recoverability of the adaptive NST based algorithms relies quite heavily on the values of $s_0$. Moreover, an optimal initial sparsity value $s_0$ for Gaussian sparse signals and Bernoulli sparse signals are different. Specifically, as indicated in Figure~\ref{F:RecoveryFrequencyAdaptive:GaussianVectors}, Gaussian sparse signals require smaller initial $s_0$ for better performance. On the other hand, for Bernoulli sparse signals, larger $s_0$ gives rise to better performance. The other three adaptive NST based algorithms all possess the similar principle. The empirical optimal values of $s_0$ for different NST based algorithms are summarized in Table~\ref{T:NearOptimalInitialization}.

\begin{table}
\centering
\begin{tabular}{|c||c|c|}
  \hline
  Adaptive NST algorithms & Gaussian sparse signals & Bernoulli sparse signals \\
  \hline
  Adaptive NST+HT & $0.2s \le s_0 \le 0.4s$ & $s_0 \approx 0.9s$ \\
  \hline
  Adaptive NST+HT+FB & $0.2s \le s_0 \le 0.4s$ & $s_0 \approx 0.9s$ \\
  \hline
  Adaptive NST+HT+subFB & $0.2s \le s_0 \le 0.4s$ & $s_0 \approx 0.9s$ \\
  \hline
  Adaptive NST+stretchedHT & $0.4s \le s_0 \le 0.5s$ & $s_0 \approx 0.9s$ \\
  \hline
\end{tabular}
\vspace{6pt}
\caption{The near optimal values of $s_0$ for various adaptive NST based algorithms.}
\label{T:NearOptimalInitialization}
\end{table}

\begin{figure}
\centering
\subfigure[Nonadaptive NST based algorithms]{
\label{F:RecoveryFrequencyNST_GaussianGaussian:AllNonAdaptive}
\includegraphics[width=0.485\columnwidth]{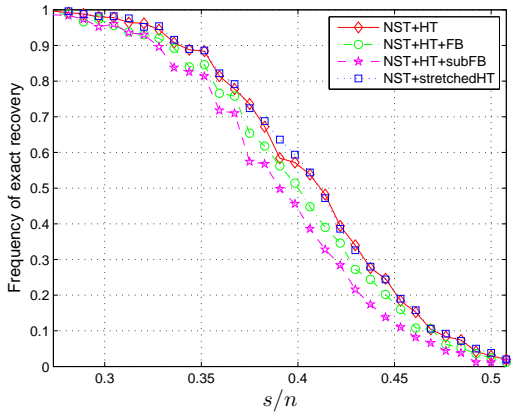}}
\subfigure[Adaptive NST based algorithms]{
\label{F:RecoveryFrequencyNST_GaussianGaussian:AllAdaptive}
\includegraphics[width=0.485\columnwidth]{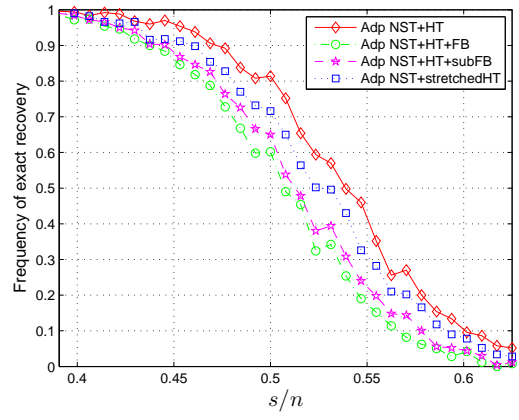}}
\caption{Plots of the frequency of exact recovery of Gaussian sparse vectors as a function of $s/n$ for (adaptive) NST based algorithms. Within the nonadaptive group, NST+HT and NST+stretchedHT outperform the other two nonadaptive algorithms slightly. Within the adaptive group, adaptive NST+HT is the best choice.}
\label{F:RecoveryFrequencyNST_GaussianGaussian}
\end{figure}

Figures~\ref{F:RecoveryFrequencyNST_GaussianGaussian} demonstrates the comparison of the recoverability of the (adaptive) NST group for recovering Gaussian sparse vectors. Figure~\ref{F:RecoveryFrequencyNST_GaussianGaussian:AllNonAdaptive} compares the nonadaptive NST based algorithms, while Figure~\ref{F:RecoveryFrequencyNST_GaussianGaussian:AllAdaptive} shows the comparison among the adaptive NST based algorithms. We comment that horizontal axis is restricted in the transition range. Consequently, even though there are some differentiations among the different algorithms, their performances are very similar. Otherwise, if we plot the overall horizontal axis, the different curves would be non-distinguishable. Similar results hold for Bernoulli sparse vectors. We decide not to show the very similar plots within the (adaptive) NST based algorithms.

\section*{Acknowledgment}
The authors would like to thank Dr. Haizhang Zhang for useful and inspiring discussions throughout the project.


\bibliographystyle{elsarticle-num}
\bibliography{Bib_NST}







\end{document}